\documentclass[aps,prl,twocolumn,superscriptaddress,floatfix]{revtex4-1}
\usepackage{graphicx,graphics}
\usepackage{dcolumn}
\usepackage{amsmath,amssymb,amsfonts}
\usepackage{latexsym,verbatim}
\usepackage{bm}
\usepackage{color}
\usepackage{ulem}
\usepackage{braket}
\usepackage[percent]{overpic}
\usepackage[breaklinks=true,colorlinks,citecolor=blue,linkcolor=blue,urlcolor=blue]{hyperref}

\begin{document}
\title{Cavity QED of Strongly Correlated Electron Systems:\\ A No-go Theorem for Photon Condensation}
\author{G.M. Andolina}
\affiliation{NEST, Scuola Normale Superiore, I-56126 Pisa,~Italy}
\affiliation{Istituto Italiano di Tecnologia, Graphene Labs, Via Morego 30, I-16163 Genova,~Italy}
\author{F.M.D. Pellegrino}
\affiliation{Dipartimento di Fisica e Astronomia ``Ettore Majorana'', Universit\`a di Catania, Via S. Sofia 64, I-95123 Catania,~Italy}
\affiliation{INFN, Sez.~Catania, I-95123 Catania,~Italy}
\author{V. Giovannetti}
\affiliation{NEST, Scuola Normale Superiore and Istituto Nanoscienze-CNR, I-56126 Pisa,~Italy}
\author{A.H. MacDonald}
\affiliation{Department of Physics, The University of Texas at Austin, Austin, TX 78712, USA}
\author{M. Polini}
\affiliation{Istituto Italiano di Tecnologia, Graphene Labs, Via Morego 30, I-16163 Genova,~Italy}
\begin{abstract}
In spite of decades of work it has remained unclear whether or not superradiant quantum phases,
referred to here as photon condensates, can occur in equilibrium.  
In this Letter, we first show that when a non-relativistic quantum many-body system is coupled to a cavity field,
gauge invariance forbids photon condensation.  
We then present a microscopic theory of the cavity quantum electrodynamics of an extended Falicov-Kimball model, showing that, in agreement with the general theorem, its insulating ferroelectric and exciton condensate phases are not altered by the cavity and do not support photon condensation.  
\end{abstract}
\maketitle

{\it Introduction.}---Superradiance~\cite{dicke_pr_1954,gross_pr_1982,cong_josaB_2016,kockum_naturereviewsphysics_2019,kirton19} 
refers to the coherent spontaneous radiation process that occurs in a dense gas when a radiation field mode
mediates long-range inter-molecule interactions. Superradiance was observed  first more than 40 years ago in optically pumped gases~\cite{gross_pr_1982,cong_josaB_2016} and has recently been identified
in optically pumped electron systems in a semiconductor quantum well placed in a perpendicular magnetic field~\cite{noe_natphys_2012}.
In 1973 Hepp and Lieb~\cite{hepp_lieb} and Wang and Hioe~\cite{wang_pra_1973} independently pointed out
that for sufficiently strong light-matter coupling  the Dicke model, often used to describe superradiance in optical 
cavities, has a finite temperature second-order {\it equilibrium} phase transition between normal and superradiant states. 
To the best of our knowledge, this phase transition has never been observed~\cite{ETH}. In the superradiant phase the ground state contains a macroscopically large number of coherent 
photons, {\it i.e.}~$\braket{\hat{a}}\neq 0$, where $\hat{a}$ ($\hat{a}^\dagger$) destroys (creates) a cavity photon.
To avoid confusion with the phenomenon discussed in the original work by Dicke~\cite{dicke_pr_1954},
we refer to the equilibrium superradiant phase as a photon condensate.

Theoretical work on photon condensation has an interesting and tortured history.  
Early on it was shown that photon condensation is robust against the addition of counter-rotating 
terms~\cite{carmichael_physlett_1973} neglected in Refs.~\cite{hepp_lieb,wang_pra_1973}.   
Soon after, however, Rza\.{z}ewski~et al.~\cite{rzazewski_prl_1975} pointed out that
addition of a neglected term related to the Thomas-Reiche-Kuhn (TRK) sum rule~\cite{Sakurai,Tufarelli15}  and proportional to $(\hat{a}+\hat{a}^\dagger)^2$ destroys the photon condensate.  These quadratic terms are naturally generated by applying minimal coupling 
$\hat{\bm p}\to \hat{\bm p}-q {\bm A}/c$ to the electron kinetic energy $\hat{\bm p}^2/(2m)$.
More recent research has focused on ground state properties. 
The quantum chaotic and entanglement properties of the Dicke model photon condensate 
were studied in Refs.~\cite{emary_brandes,buzek_prl_2005}. The authors of Ref.~\cite{rzazewski_prl_2006} criticized these studies however, pointing again to the importance of the quadratic term. 
The no-go theorem for photon condensation was revisited in Ref.~\cite{nataf_naturecommun_2010}, where it was
claimed that it can be bypassed in a circuit quantum electrodynamics (QED) system with Cooper pair boxes 
capacitively coupled to a resonator. 
Soon after, however, Ref.~\cite{viehmann_prl_2011} showed that the no-go theorem for 
cavity QED applies to circuit QED as well. The claims of Ref.~\cite{viehmann_prl_2011} were then criticized in Ref.~\cite{ciuti_prl_2012}. (See also subsequent discussions~\cite{Jaako16,Bamba16} on light-matter interactions in circuit QED.) Later it was argued~\cite{hagenmuller_prl_2012} that the linear band dispersion of graphene 
provides a route to bypass the no-go theorem, and that photon condensation could occur in graphene in the integer quantum Hall regime. This claim was later countered in Refs.~\cite{chirolli_prl_2012,pellegrino_prb_2014},
where it was shown that a dynamically generated quadratic term again forbids photon condensation.

Recent experimental progress has created opportunities to study light matter interactions in new
regimes in which direct electron-electron interactions play a prominent role.
For example~\cite{pellegrino_natcom_2016} two-dimensional (2D) electron systems 
can be embedded in cavities or exposed to the radiation field of 
metamaterials, making it possible to study strong light-matter interactions in the quantum Hall regime~\cite{scalari_science_2012,muravev_prb_2013,smolka_science_2014,ravets_prl_2018,Paravicini-Bagliani_natphys_2019,knuppel_arxiv_2019}. Other emerging possibilities
include cavity QED with quasi-2D electron systems 
that exhibit exciton condensation, superconductivity, magnetism, or Mott insulating states.
This Letter is motivated by interest in strong light-matter interactions in these new regimes and 
by fundamental confusion on when, if ever, photon condensation is allowed.  
We present a no-go theorem for photon condensation that is valid for generic non-relativistic 
interacting electrons at $T=0$. This result generalizes to interacting systems existing no-go theorems for photon 
condensation in two-level~\cite{nataf_naturecommun_2010,rzazewski_prl_1975} and 
multi-level~\cite{viehmann_prl_2011} Dicke models.  
We then present a theory of cavity QED of an extended Falikov-Kimball model~\cite{EFKM}, which, in the absence of the 
cavity, has insulating ferroelectric and exciton condensate phases. 
We show through explicit microscopic calculations how the theorem is satisfied in this 
particular strongly correlated electron model.

{\it Gauge invariance excludes photon condensation.}---We consider a system of $N$ electrons of mass $m_{i}$ described by a non-relativistic many-body Hamiltonian of the form
\begin{equation}\label{eq:Hamiltonian}
\hat{\cal H}=\sum_{i=1}^N  \left[\frac{\hat{ {\bm p}}_i^2}{2m_i}+V(\hat{\bm r}_i)\right] + \frac{1}{2}\sum_{i\neq j} v(\hat{\bm r}_{i} - \hat{\bm r}_{j})~.
\end{equation}
Here, $V({\bm r})$ is a generic function of position and $v({\bm r})$ is a generic (non-retarded) two-body interaction, which need not even be spherically symmetric. 
In a solid $V({\bm r})$ is the one-body the crystal potential. Below we first exclude the possibility of 
a continuous transition to a condensed state, and then use this insight to exclude first-order transitions.
For future reference, we denote by $\ket{\psi_m}$ and $E_{m}$ the {\it exact} eigenstates and eigenvalues of 
$\hat{\cal H}$~\cite{Giuliani_and_Vignale,Pines_and_Nozieres}, with $\ket{\psi_{0}}$ and $E_{0}$ denoting the 
ground state and ground-state energy, respectively.

We treat the cavity e.m. field in a quantum fashion, via a uniform quantum field $\hat{{\bm A}}$ corresponding to only one mode~\cite{DeBernardis18,DeBernardis18b,Jaako16,Bamba16,Tufarelli15,kirton19,hepp_lieb,wang_pra_1973,carmichael_physlett_1973,rzazewski_prl_1975,emary_brandes,nataf_naturecommun_2010,viehmann_prl_2011,ciuti_prl_2012,hagenmuller_prl_2012,chirolli_prl_2012,pellegrino_prb_2014,keeling_jpcm_2007}, i.e.~$\hat{{\bm A}}=A_0 {\bm u}(\hat{a} + \hat{a}^\dagger)$, where ${\bm u}$ is the polarization vector, $A_{0} =\sqrt{2\pi \hbar c^2/(V \omega_{\rm c}\epsilon_{\rm r})}$, $V$ is the volume of the cavity, $\epsilon_{\rm r}$ is  its relative dielectric constant, and the photon Hamiltonian $\hat{\cal H}_{\rm ph} = \hbar\omega_{\rm c}\hat{a}^\dagger\hat{a}$,
where $\omega_{\rm c}$ is the cavity frequency. The full Hamiltonian, including light-matter interactions in the Coulomb gauge~\cite{keeling_jpcm_2007,Vukics12,Stokes19} is:
\begin{eqnarray}\label{eqHtot}
\hat{\cal H}_{{\bm A}_{0}}&=&\hat{\cal H}+ \hbar \omega_{\rm c} \hat{a}^\dagger \hat{a}+\sum_{i=1}^{N} \frac{e}{m_i c} \hat{ {\bm p}}_i \cdot {\bm A}_0  (\hat{a} + \hat{a}^\dagger)\nonumber\\
&+&\sum_{i=1}^{N} \frac{e^2A^2_{0}}{2m_ic^2}(\hat{a} + \hat{a}^\dagger)^2~,
\end{eqnarray}
where ${\bm A}_{0}\equiv A_{0} {\bm u}$ and $-e<0$ is the electron charge. 
The third and fourth terms in Eq.~(\ref{eqHtot}) are often referred to respectively as the 
paramagnetic and diamagnetic contributions to the light-matter coupling Hamiltonian.
Our aim is to make general statements about  the ground state $\ket{\Psi}$ of $\hat{\cal H}_{{\bm A}_{0}}$.
For future reference we define i) the paramagnetic (number) current operator~\cite{Giuliani_and_Vignale,Pines_and_Nozieres}, 
$\hat{{\bm j}}_{\rm p}\equiv (c/e)\delta \hat{\cal H}_{{\bm A}_{0}}/\delta {\bm A}_{0}|_{{\bm A}_{0} = {\bm 0}} 
= \sum_{i=1}^{N} \hat{ {\bm p}}_i/m_i$ and ii) $\Delta \equiv \sum_{i=1}^{N} e^2 A^2_0/( 2m_{i} c^2)$.

The term proportional to $\Delta$ in Eq.~(\ref{eqHtot}) can be removed by performing the transformation $\hat{b} = \cosh(x) \hat{a} + \sinh(x) \hat{a}^\dag$, where $\cosh(x) = (\lambda+1)/(2\sqrt{\lambda})$ and 
$\sinh(x) = (\lambda-1)/(2\sqrt{\lambda})$ with $\lambda=\sqrt{1+4\Delta/(\hbar \omega_{\rm c})}$. The Hamiltonian (\ref{eqHtot}) becomes: $\hat{\cal H}_{{\bm A}_{0}}= \hat{\cal H}+(e/c) \hat{\bm j}_{\rm p}\cdot {\bm A}_{0}\lambda^{-1/2} (\hat{b} + \hat{b}^\dagger) + \hbar\omega_{\rm c} \lambda \hat{b}^\dagger \hat{b}$.  It can be shown (see Sec.~I of the Supplemental Material (SM)~\cite{SOM}) that in the thermodynamic limit ($N\to\infty$, $V\to \infty$ limit at fixed $N/V$), the ground state $\ket{\Psi}$ of $\hat{\cal H}_{{\bm A}_{0}}$ does not contain light-matter entanglement, i.e.~we can take $\ket{\Psi}=\ket{\psi}\ket{\Phi}$, where $\ket{\psi}$ and $\ket{\Phi}$ are matter and light wave functions.  
Using this property we see that in the thermodynamic limit
the ground state $|\Phi\rangle$ of the effective photon Hamiltonian $\bra{\psi}\hat{\cal H}_{{\bm A}_{0}}\ket{\psi}$ is a coherent state~\cite{Walls_and_Milburn,Serafini} $|\beta\rangle$ satisfying $\hat{b}\ket{\beta}=\beta\ket{\beta}$. The ground-state energy is therefore given by
\begin{equation}\label{eqHL2}
E_{\psi}(\beta)=\braket{ \psi|\hat{\cal H}|\psi} + \frac{e}{c}\braket{\psi|\hat{\bm j}_{\rm p}|\psi} \cdot {\bm A}_{0}\frac{2{\rm Re}[\beta]}{\sqrt{\lambda}} + \hbar \omega_{\rm c} \lambda |\beta|^2~.
\end{equation}
We need to minimize $E_{\psi}(\beta)$ with respect to $\beta$ and $|\psi\rangle$.
The minimization with respect to $\beta$ can be done analytically. We find that the optimal value $\bar{\beta}$ is a real number given by:
\begin{equation}\label{eq:optimal_beta}
\bar{\beta}=-\frac{1}{\hbar \omega_{\rm c}\lambda^{3/2}} \frac{e}{c}\braket{\psi|\hat{\bm j}_{\rm p}|\psi}\cdot {\bm A}_0~.
\end{equation}
We are therefore left with a {\it constrained} minimum problem for the matter degrees of freedom. Its solution must be sought among the normalized anti-symmetric states $\ket{\psi}$  which yield (\ref{eq:optimal_beta}). This is the typical scenario that can be handled with the stiffness theorem~\cite{Giuliani_and_Vignale}.

For photon condensation to occur we need $E_{\psi}(\bar{\beta})< E_{\psi_0}(0)$
or, equivalently,
\begin{eqnarray}\label{eqcond1}
\hbar\omega_{\rm c}\lambda\bar{\beta}^2>\braket{ \psi|\hat{\cal H}|\psi}-\braket{ \psi_0|\hat{\cal H}|\psi_0}~,
\end{eqnarray} 
where, because of (\ref{eq:optimal_beta}), $\ket{\psi}$ depends on $\bar{\beta}$. 
The dependence of $\braket{ \psi|\hat{\cal H}|\psi}-\braket{ \psi_0|\hat{\cal H}|\psi_0}$ on $\bar{\beta}$ can be calculated {\it exactly} up to order $\bar{\beta}^2$ by using the stiffness theorem~\cite{Giuliani_and_Vignale}. 
We find $\braket{ \psi|\hat{\cal H}|\psi}-\braket{ \psi_0|\hat{\cal H}|\psi_0}= \alpha\bar{\beta}^2/2 + {\cal O}(\bar{\beta}^3)$, where $\alpha=-1/\chi(0)>0$ and
\begin{equation}\label{eq:exact_eigenstate_chiBB}
\chi(0) \equiv -\frac{2}{\hbar^2 \omega^2_{\rm c}\lambda^{3} } \frac{e^2}{c^2} \sum_{n\neq 0}\frac{|\langle \psi_{n}|\hat{\bm j}_{\rm p}\cdot {\bm A}_{0}|\psi_{0}\rangle|^2}{E_{n}-E_{0}}< 0
\end{equation}
is proportional to the static paramagnetic current-current response function in the Lehmann representation~\cite{Giuliani_and_Vignale,Pines_and_Nozieres}.
We have used that $(e/c)\braket{\psi_0|\hat{\bm j}_{\rm p}|\psi_0}\cdot {\bm A}_{0} =0$,
as proven in Sec.~II of the SM~\cite{SOM}.
It follows that photon condensation occurs if and only if 
\begin{eqnarray}\label{eqcond3}
4\frac{e^2}{c^2}\sum_{n\neq0}\frac{|\langle \psi_{n}|\hat{\bm j}_{\rm p}\cdot {\bm A}_{0}|\psi_{0}\rangle|^2}{E_{n}-E_0} >\hbar \omega_{\rm c}+4\Delta~.
\end{eqnarray} 
However, as shown in Sec.~III of the SM~\cite{SOM}, 
\begin{equation}\label{eq:explicit_form_gauge_invariance}
\frac{e^2}{c^2}\sum_{n\neq0}\frac{|\langle \psi_{n}|\hat{\bm j}_{\rm p}\cdot {\bm A}_{0}|\psi_{0}\rangle|^2}{E_{n}-E_0} =\Delta~.
\end{equation}
Eq.~(\ref{eq:explicit_form_gauge_invariance}) is  the TRK sum rule~\cite{Sakurai} which expresses
the fact that the paramagnetic and diamagnetic contributions to the physical current-current response function 
cancel in the uniform static limit~\cite{Giuliani_and_Vignale,Pines_and_Nozieres},
as discussed more fully in Sec.~III of the SM~\cite{SOM}, i.e.~it expresses gauge invariance.  Using Eq.~(\ref{eq:explicit_form_gauge_invariance}) we can finally rewrite Eq.~\eqref{eqcond3} as $c^2 4\Delta >c^2(\hbar \omega_{\rm c}+4\Delta)$ which cannot be satisfied. We conclude that photon condensation cannot occur and that, upon minimization with respect to $|\psi\rangle$, the ground state is $|\psi_{0}\rangle$ and  $\bar{\beta}=0$.
From this analysis it is clear that first-order transitions to states with finite photon density are also
excluded, because interactions with a coherent equilibrium photon field do not lower the matter energy~\cite{first-order}. Gauge invariance excludes photon condensation for any Hamiltonian of the form (\ref{eqHtot}). This is the first important result of this Letter.

{\it Cavity QED of an extended Falikov-Kimball model.}---We now illustrate how this general conclusion applies to a specific 
properly gauge invariant model of strongly correlated electrons in a cavity.  
We consider spinless electrons in a one-dimensional (1D) inversion-symmetric crystal with $N$ sites, each with one atom with two 
atomic orbitals of opposite parity (${\rm s}$ and ${\rm p}$). 
When this lattice model is augmented by the addition of on-site repulsive electron-electron interactions, it is often 
referred to as an extended Falikov-Kimball (EFK) model~\cite{EFKM}. 
The EFK model has been used to discuss exciton condensation~\cite{excitonic_insulators} and electronic ferroelectricity~\cite{portengen_prl_1996,batista_prl_2002}.  The coupling of cavity 
photons to the matter degrees of freedom of a 1D EFK model can be 
described~\cite{kohn_pr_1964,shastry_prl_1990,millis_prb_1990,fye_prb_1991} by 
employing a Peierls substitution in the site representation 
with a uniform linearly-polarized vector potential of amplitude $A_{0}$, as detailed in Sec.~IV of the SM~\cite{SOM}.
We obtain
\begin{eqnarray}\label{eq:quadratic_Hamiltonian_final}
\hat{\cal H}_{{\bm A}_{0}}&=&\hat{\cal H}_{0}+\hat{\cal H}_{\rm ee} + \hbar\omega_{\rm c}\hat{a}^\dagger\hat{a}+ \frac{g_{0}}{\sqrt{N}}\frac{\hbar}{a}\hat{j}_{\rm p}(\hat{a} + \hat{a}^\dagger)\nonumber\\
&-&\frac{g^2_{0}}{2N}\hat{\cal T} (\hat{a} + \hat{a}^\dagger)^2~,
\end{eqnarray}
where $\hat{\cal H}_{0} = \sum_{k,\alpha,\beta}\hat{c}^\dagger_{k,\alpha}H_{\alpha\beta}(k)\hat{c}_{k,\beta}$ is the band Hamiltonian,
\begin{equation}\label{eq:band_energy}
H_{\alpha\beta}(k) = \begin{pmatrix}
E_{\rm s} -2 t_{\rm s} \cos(ka) & 2 i \tilde{t} \sin(ka)\\
-2 i \tilde{t} \sin(ka) &  E_{\rm p} + 2 t_{\rm p} \cos(ka)
\end{pmatrix}~,
\end{equation}
and the Hubbard interaction term
\begin{equation}\label{eq:Hubbard}
\hat{\cal H}_{\rm ee} =U \sum_{j=1}^{N} \hat{c}^\dag_{j, {\rm s}} \hat{c}_{j, {\rm s}} \hat{c}^\dag_{j, {\rm p}} \hat{c}_{j,{\rm p}}~.
\end{equation}
In Eq.~(\ref{eq:quadratic_Hamiltonian_final}), $\hat{j}_{\rm p} = \sum_{k,\alpha,\beta}\hat{c}^\dagger_{k,\alpha}j_{\alpha\beta}(k)\hat{c}_{k,\beta}$ with $j_{\alpha\beta}(k) \equiv \hbar^{-1}\partial H_{\alpha\beta}(k)/\partial k$ is the paramagnetic number current operator, and $\hat{\cal T} = \sum_{k,\alpha,\beta}\hat{c}^\dagger_{k,\alpha}{\cal T}_{\alpha\beta}(k)\hat{c}_{k,\beta}$ with 
${\cal T}_{\alpha\beta}(k) \equiv -a^{-2}\partial^2 H_{\alpha\beta}(k)/\partial k^2$ is the diamagnetic operator. 
In Eq.~(\ref{eq:band_energy}), $E_{\rm s}$ and $E_{\rm p}$ are on-site energies for the ${\rm s}$ and ${\rm p}$ orbitals, $t_{\rm s}\in \mathbb{R}$ and $t_{\rm p}\in \mathbb{R}$ are hopping parameters, 
and $\tilde{t}\in \mathbb{R}$ is the inter-band hopping parameter.
At the single-particle level (i.e.~for $U=0$), $\tilde{t}$ is the only term responsible for inter-band transitions due to light. All sums over the wave number $k$ are carried out in the 1D Brillouin zone and become integrals in the thermodynamic limit with the usual rule $N^{-1}\sum_{k}\to a\int_{-\pi/a}^{+\pi/a}dk/(2\pi)$, where $a$ is the lattice constant. In these equations the Greek labels
take values $\alpha,\beta={\rm s}, {\rm p}$. The momentum-space and site representations for field operators are 
linked by the usual relationship $\hat{c}^\dagger_{j,\alpha} = N^{-1/2}\sum_{k} \hat{c}^\dagger_{k,\alpha} e^{-i k j a}$. 
The dimensionless light-matter coupling constant in Eq.~(\ref{eq:quadratic_Hamiltonian_final}) is defined by 
$g \equiv e a A_{0}/(\hbar c) = g_{0}/\sqrt{N}$, where $g_{0} \equiv \sqrt{2\pi e^2/(\hbar v_{0} \omega_{\rm c}\epsilon_{\rm r})}$
and $v_{0}=V/N$ is the cavity volume per site. 

We emphasize that the  operators $\hat{j}_{\rm p}$ and $\hat{\cal T}$ describing light-matter interactions are completely determined 
by the matrix elements $H_{\alpha\beta}(k)$ of the band Hamiltonian. This  property is crucial to have a 
properly gauge-invariant model~\cite{f-sum-rule} and must be a general feature of {\it any} strongly correlated lattice model 
coupled to cavity photons.

In the limit $g_{0}\to 0$, the model reduces to a 1D EFK model~\cite{EFKM,portengen_prl_1996,batista_prl_2002}. In the limit $ka\to 0$ and $U=0$, Eq.~(\ref{eq:quadratic_Hamiltonian_final}) reduces to the Dicke model, augmented by the addition of a term proportional to $\sum_{k,\alpha,\beta}\hat{c}^\dagger_{k,\alpha}\sigma^{(z)}_{\alpha\beta}\hat{c}_{k,\beta}(\hat{a}+\hat{a}^\dagger)^2$~\cite{distefano_arXiv_2018,DeBernardis18,DeBernardis18b}, where $\sigma^{(z)}_{\alpha\beta}$ are the matrix elements of the corresponding $2\times2$ Pauli matrix. For non-interacting systems, the diamagnetic term prevents photon condensation from occurring 
in the thermodynamic limit~\cite{rzazewski_prl_1975,nataf_naturecommun_2010}. We now show that interactions 
do not help.  $\hat{\cal H}_{{\bm A}_{0}}$ does not support photon condensation.

To make progress in analyzing the interacting problem 
we treat the Hubbard term using an unrestricted Hartree-Fock (HF) 
approximation~\cite{Giuliani_and_Vignale,verges}. As detailed in Sec.~V of the SM~\cite{SOM} we arrive at 
\begin{eqnarray}\label{eq:VMF*}
\hat{\cal H}^{({\rm HF})}_{\rm ee} &=& -U \frac{\cal M}{2}\sum_{k} 
(\hat{c}^\dag_{k,{\rm p}} \hat{c}_{k,{\rm p}} - \hat{c}^\dag_{k,{\rm s}} \hat{c}_{k,{\rm s}})\nonumber\\
&-&U\sum_{k}(
{\cal I} \hat{c}^\dag_{k,{\rm s}} \hat{c}_{k,{\rm p}}  
+ {\cal I}^\ast  \hat{c}^\dag_{k,{\rm p}} \hat{c}_{k,{\rm s}}) +U \frac{n_0}{2}\sum_{k,\alpha} 
\hat{n}_{k,\alpha} \nonumber\\
&+&U N \left( \frac{{\cal M}^2-n^2_{0}}{4} + |{\cal I}|^2 \right)~.
\end{eqnarray}
In Eq.~(\ref{eq:VMF*}) we have introduced the following self-consistent fields: i) the electronic polarization
\begin{equation}
{\cal M} \equiv
\frac{1}{N} \sum_{k} ( \langle  \hat{c}^\dag_{k,{\rm p}} \hat{c}_{k,{\rm p}} \rangle - \langle  \hat{c}^\dag_{k,{\rm s}} \hat{c}_{k,{\rm s}} \rangle)~,
\end{equation}
ii) the complex excitonic order parameter
\begin{equation}
{\cal I} \equiv 
\frac{1}{N} \sum_{k} \langle  \hat{c}^\dag_{k,{\rm p}} \hat{c}_{k,{\rm s}} \rangle~,
\end{equation}
and iii) the number of electrons per site $n_0 \equiv  N^{-1} \sum_{k,\alpha}\langle  \hat{n}_{k,\alpha} \rangle$, where $\hat{n}_{k,\alpha} \equiv \hat{c}^\dag_{k,\alpha} \hat{c}_{k,\alpha}$.
The term proportional to $n_{0}/2$ in Eq.~(\ref{eq:VMF*}) acts as a renormalization of the chemical potential in the grand-canonical Hamiltonian and can be discarded in this study since we study the phase diagram only at half filling and $n_{0}=1$ in all phases. 

In order to reduce the number of free parameters in the problem, from now on we enforce particle-hole symmetry in the bare band Hamiltonian $\hat{\cal H}_{0}$ by setting $E_{\rm s} \equiv -E_{\rm p} = -E_{\rm g}/2$ and $t_{\rm s}\equiv t_{\rm p}=t$ (with $|t|>E_{\rm g}/4$, see Fig.~\ref{fig:S1}). In order to find the ground state of the Hamiltonian (\ref{eq:quadratic_Hamiltonian_final}) with Hubbard interactions treated as in Eq.~(\ref{eq:VMF*}), we follow the same steps outlined in the proof of the no-go theorem above.  We seek a ground state of the unentangled form $\ket{\Psi}=\ket{\psi}\ket{\Phi}$. After removing the term proportional to $(\hat{a}+\hat{a}^\dagger)^2$, one finds that $\ket{\Phi}$ must be a coherent state $\ket{\bar{\beta}}$ with $\bar{\beta}=-g_{0}{\cal J}\sqrt{N}/(\lambda^{3/2}\hbar\omega_{\rm c})$. 
(We remind the reader that the photon condensate order parameter is $\braket{\bar{\beta}|\hat{a}|\bar{\beta}}/\sqrt{N}=
\langle\bar{\beta}|\cosh(x)\hat{b}-\sinh(x)\hat{b}^\dagger|\bar{\beta}\rangle/\sqrt{N}=\bar{\beta}/\sqrt{N\lambda}$. See Sec.~VI of the SM~\cite{SOM}.) Here, ${\cal J}\equiv \hbar\bra{\psi}\hat{j}_{\rm p}\ket{\psi}/(a N)$, $\lambda$ has the same expression as in the proof of the no-go theorem with $\Delta=-g^2_{0}{\cal T}/2$, and ${\cal T}\equiv \bra{\psi}\hat{\cal T}\ket{\psi}$. Note that both ${\cal J}$ and ${\cal T}$ have units of energy and are finite in the $N\to \infty$ limit.

The resulting effective Hamiltonian for the matter degrees of freedom, i.e.~$\bra{\bar{\beta}}\hat{\cal H}_{{\bm A}_{0}}\ket{\bar{\beta}}$, can be diagonalized exactly since, after the HF decoupling, it is quadratic in the fermionic operators $\hat{c}_{k,\alpha}$, $\hat{c}^\dagger_{k,\alpha}$. To this end, it is sufficient to introduce the Bogoliubov operators 
$\hat{\gamma}^\dagger_{k,-}=u_{k} \hat{c}^\dagger_{k,{\rm s}}+v_{k} \hat{c}^\dagger_{k,{\rm p}}$ and 
$\hat{\gamma}^\dagger_{k,+}=v^*_{k} \hat{c}^\dagger_{k,{\rm s}}-u^*_{k} \hat{c}^\dagger_{k,{\rm p}}$, 
where the quantities $u_{k}$ and $v_{k}$ depend on the parameters of the bare Hamiltonian $\hat{\cal H}_{0}$, on the Hubbard parameter $U$, on the light-matter coupling constant $g_{0}$, and on the quantities ${\cal I}$, ${\cal M}$, ${\cal J}$, and ${\cal T}$. The ground state $\ket{\psi}=\prod_{k} \hat{\gamma}^\dagger_{k,-}\ket{\rm vac}$ can be written in a BCS-like fashion, 
\begin{equation}\label{eq:BCS}
\ket{\psi}=\prod_{k} \big [  u_{k} +v_{k} \hat{c}^\dagger_{k,{\rm p}} \hat{c}_{k,{\rm s}} \big]  \ket{\emptyset}~,
\end{equation}
where $\ket{\emptyset}=\prod_{k} \hat{c}^\dagger_{k,{\rm s}}\ket{\rm vac}$ and $\ket{\rm vac}$ is the state with no electrons. The 
final ingredients which are needed are the expressions for the quantities ${\cal M}$, ${\cal I}$, ${\cal J}$, and ${\cal T}$ in terms of $u_{k},v_{k}$: ${\cal M}=N^{-1}\sum_{k}(|v_{k}|^2-|u_{k}|^2)$, ${\cal I}=N^{-1}\sum_{k} v^{*}_{k}u_{k}$, ${\cal J} =2N^{-1}\sum_{k}[- t\sin(ka)(|v_{k}|^2-|u_{k}|^2)-2\tilde{t}\cos(ka){\rm Im}(u^\ast_{k}v_{k})]$, and ${\cal T}=2N^{-1}\sum_{k}[ t\cos(ka)   \big(|v_{k}|^2-|u_{k}|^2\big)-2\tilde{t}\sin(ka) {\rm Im}(u^*_{k}v_{k})]$.  The technical details of this calculation are 
summarized in Sec.~VI of the SM~\cite{SOM}.
 
\begin{figure}[t] 
\centering
\begin{overpic}[width=0.49\columnwidth]{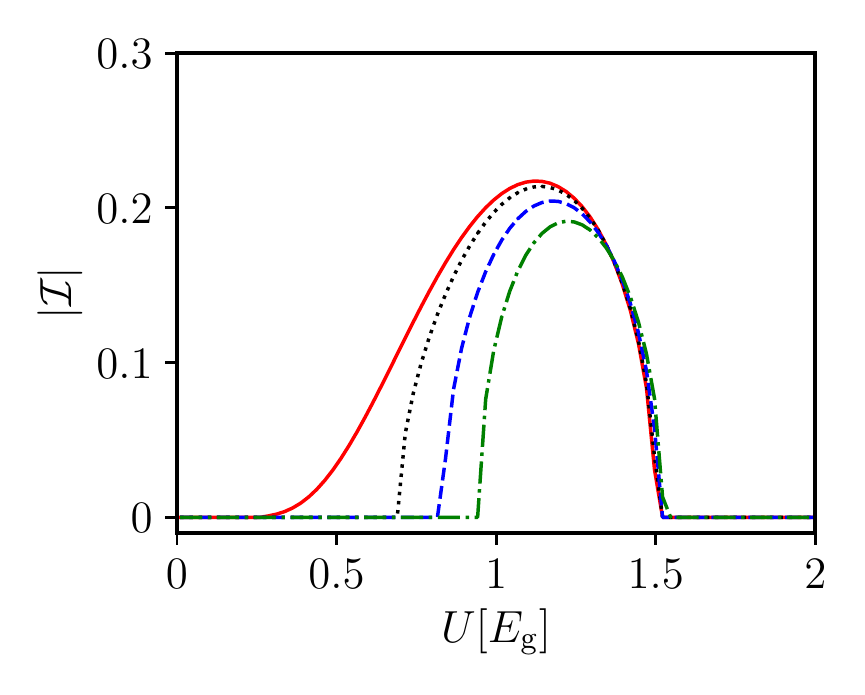}\put(4,80){(a)}\end{overpic}
\begin{overpic}[width=0.49\columnwidth]{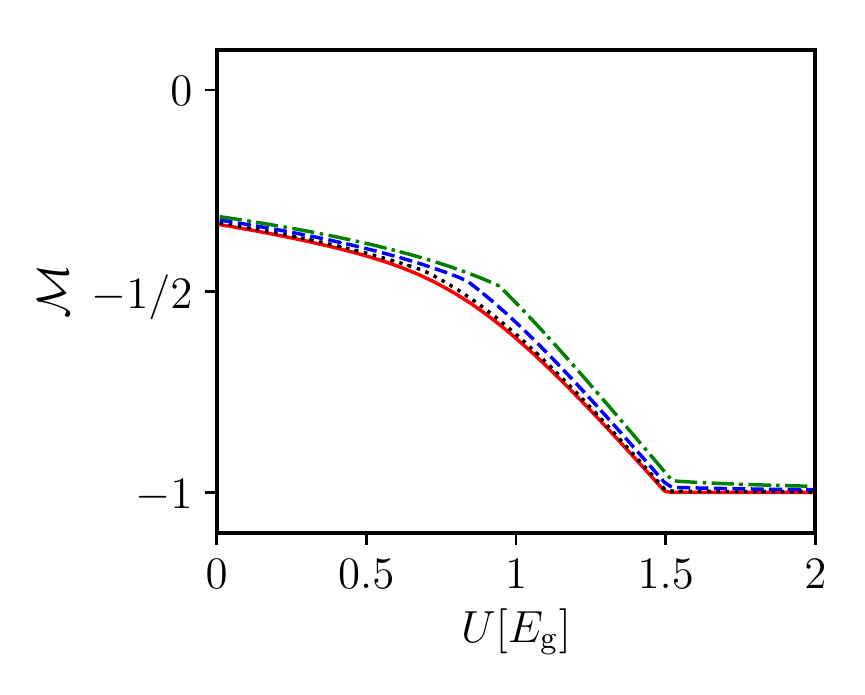}\put(4,80){(b)}\end{overpic}
\begin{overpic}[width=0.49\columnwidth]{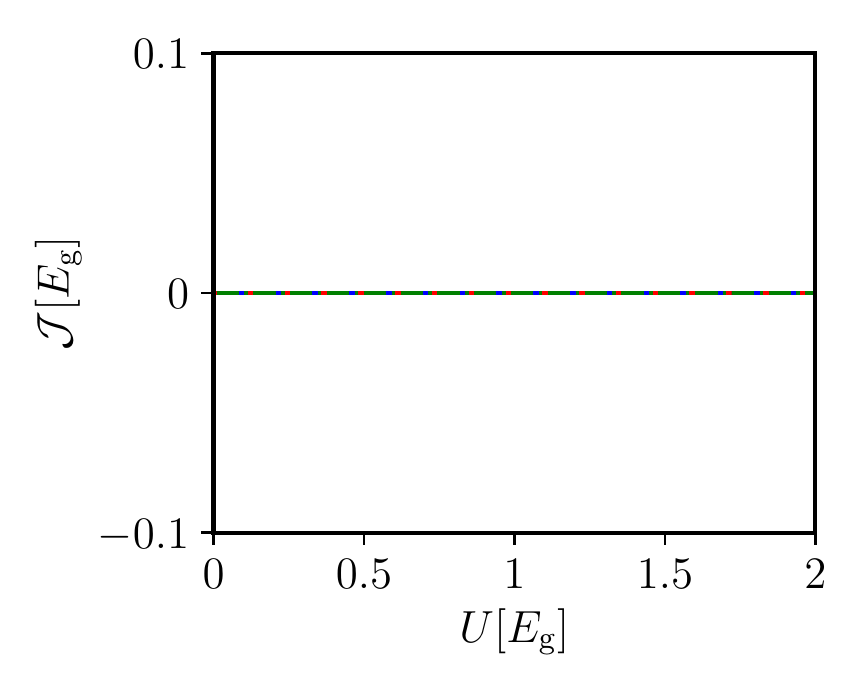}\put(4,80){(c)}\end{overpic}
\begin{overpic}[width=0.49\columnwidth]{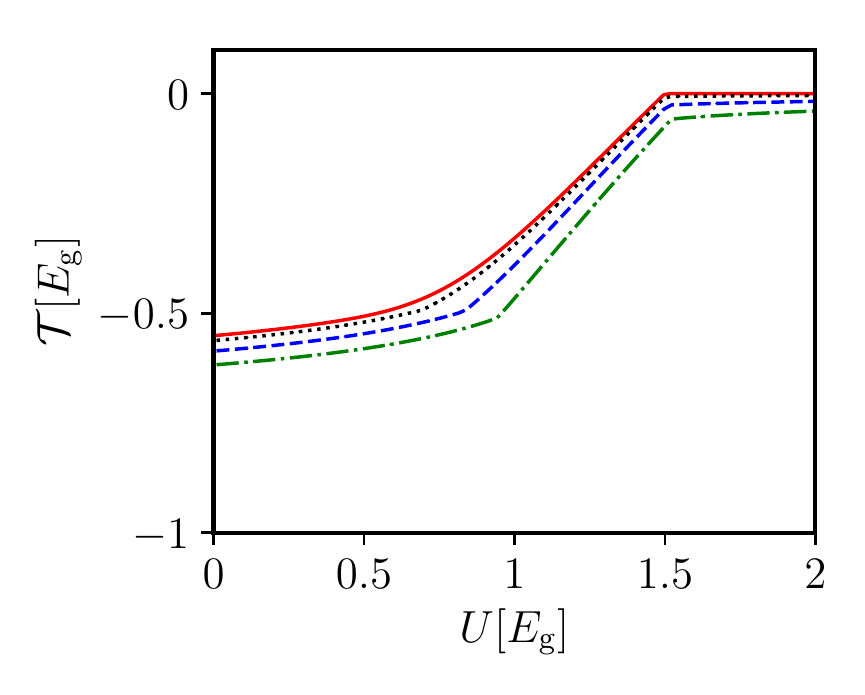}\put(4,80){(d)}\end{overpic}
  \caption{(Color online) Panel (a) The excitonic order parameter $|{\cal I}|$ is plotted as a function of $U$ (in units of $E_{\rm g}$). Numerical results have been obtained by setting $t=0.5~E_{\rm g}$ and $\hbar \omega_{\rm c}=E_{\rm g}$. Different curves correspond to different values of $\tilde{t}$. Red solid line: $\tilde{t}=10^{-4}~E_{\rm g}$. Black dotted line: $\tilde{t}=0.05~E_{\rm g}$. Blue dashed line: $\tilde{t}=0.1~E_{\rm g}$. Green dash-dotted line: $\tilde{t}=0.15~E_{\rm g}$. Note that for $\tilde{t}\neq 0$, $|{\cal I}|\neq 0$ for $U_{{\rm c}1}<U<U_{{\rm c}2}$. Panel (b) Same as in panel (a) but for the electronic polarization ${\cal M}$. Panel (c) Same as in other panels but for ${\cal J}$. Note that ${\cal J}=0$ for all values of $\tilde{t}$ and $U/E_{\rm g}$. This implies $\bar{\beta}=0$ and therefore no photon condensation. Panel (d) Same as in other panels but for ${\cal T}$ (in units of $E_{\rm g}$).\label{fig:one}}
\end{figure}

The quantities ${\cal I}$, ${\cal M}$, ${\cal J}$, and ${\cal T}$ can be determined
 by solving this nonlinear system of equations. A typical solution is shown in Fig.~\ref{fig:one}. We have found that all
 observables are independent of $g_{0}$.   
 In other words, in the thermodynamic limit the ground state is given by Eq.~(\ref{eq:BCS}) with $u_{k}$ and $v_{k}$ evaluated at $g_{0}=0$, in agreement with the general theorem proven above. 
The self-consistent solutions always have  ${\cal J}=0$ (i.e.~$\bar{\beta}=0$), as clearly seen in Fig.~\ref{fig:one}(c), and therefore display no photon condensation but may have finite polarization and exciton order parameters.  This is the second important result of this Letter. At $\tilde{t}=0$ the HF ground state has a single transition at $U=U_{\rm XC}$. For $0<U<U_{\rm XC}$ the ground state is an exciton condensate with spontaneous coherence between
${\rm s}$ and ${\rm p}$ bands~\cite{portengen_prl_1996,batista_prl_2002} which are not hybridized when $U=0$.
The ordered state appears on the small $U$ side of the transition because interactions favor orbital
polarization over coherence.  
The value of $U_{\rm XC}$ can be determined analytically
as detailed in Sec.~VIII of the SM~\cite{SOM}.  We find,
in agreement with earlier work~\cite{kocharian_prb_1996,ejima_prl_2014}, that $U_{\rm XC}=8 t^2/E_{\rm g}- E_{\rm g}/2$.

In the limit $\tilde{t}=0$, $\hat{\cal H}_{{\bm A}_{0}}$ separately conserves the number of electrons with band 
indices $\alpha={\rm s}, {\rm p}$, and has a global $U(1)$ symmetry associated with
the arbitrariness of the relative phase between s and p electrons~\cite{EFKM}. The HF ground state breaks this symmetry.
For $\tilde{t}\ \ne 0$ the $U(1)$ symmetry is reduced to a discrete $Z_{2}$ symmetry 
reflecting the invariance of the Hamiltonian under spatial inversion. 
This symmetry is broken for $U_{{\rm c}1}(\tilde{t}) < U < U_{{\rm c}2}(\tilde{t})$.
Note that $\lim_{\tilde{t}\to 0}  U_{{\rm c}2}(\tilde{t}) = U_{\rm XC}$.  
Corrections to  $U_{{\rm c}2}(0)$ can be found perturbatively for $\tilde{t}/t\ll 1$ and 
are of ${\cal O}(\tilde{t}^2)$ (see Sec.~VIII of the SM~\cite{SOM}). 
For $0<U<U_{{\rm c}1}(\tilde{t})$ inversion symmetry is unbroken and ${\cal I}=0$. 
For $U >  U_{{\rm c}1}(\tilde{t})$ the ground state is an insulating ferroelectric that breaks the $Z_{2}$ symmetry (see Sec.~IX of the SM~\cite{SOM}). 
The dependence of $U_{{\rm c}1}$ on $\tilde{t}$ in non-analytical and can be extracted 
asymptotically for $\tilde{t}/t\ll 1$. 
We find that $U_{\rm c1}(\tilde{t}) \to \pi (4 t^2-E_{\rm g}^2/4)^{1/2}/|\ln(\tilde{t}/t)|$ (see Sec.~VIII of the SM~\cite{SOM}).

In summary, we have presented a no-go theorem for photon condensation that applies to all quantum many-body Hamiltonians of 
the form (\ref{eq:Hamiltonian}), greatly extending previous no-go theorems for Dicke-type Hamiltonians~\cite{nataf_naturecommun_2010,viehmann_prl_2011}. Since the proof is non-perturbative in the strength of electron-electron interactions, our arguments against photon condensation apply to all lattice models of strongly correlated electron systems that can be derived from Eq.~(\ref{eq:Hamiltonian}). We have then explained how the theorem manifests in practice, 
presenting a theory of cavity QED of a 1D model that supports insulating ferroeletric and exciton condensate phases. 
We have shown that these electronic orders are never entwined with photon condensation~\cite{mazza_prl_2019}. In the future, it will be interesting to study the role of spatially-varying multimode cavity fields and their interplay with retarded interactions~\cite{Schlawin_prl_2019,curtis_prl_2019}, or strong magnetic fields~\cite{GMP}. Our work emphasizes that theoretical models of interacting light-matter systems 
must retain precise gauge invariance, which is often lost when the matter system
is projected onto a low-energy model.  

{\it Acknowledgements.}---A.H.M. was supported by Army Research Office (ARO) Grant \# W911NF-17-1-0312 (MURI),
and by Welch foundation Grant TBF1473.
It is a great pleasure to thank M.I. Katsnelson and F.H.L. Koppens for useful discussions.

\clearpage 
\setcounter{section}{0}
\setcounter{equation}{0}%
\setcounter{figure}{0}%
\setcounter{table}{0}%

\setcounter{page}{1}

\renewcommand{\thetable}{S\arabic{table}}
\renewcommand{\theequation}{S\arabic{equation}}
\renewcommand{\thefigure}{S\arabic{figure}}

\renewcommand{\bibnumfmt}[1]{[S#1]}
\renewcommand{\citenumfont}[1]{S#1}
\onecolumngrid

\begin{center}
\textbf{\Large Supplemental Material for ``Cavity QED of Strongly Correlated Electron Systems: A No-go Theorem for Photon Condensation''}
\bigskip

G.M. Andolina,$^{1,\,2}$
F.M.D. Pellegrino,$^{3\,,4}$
V. Giovannetti,$^5$
A.H. MacDonald,$^{6}$ and
M. Polini$^1$

\bigskip

$^1$\!{\it Istituto Italiano di Tecnologia, Graphene Labs, Via Morego 30, I-16163 Genova,~Italy}

$^2$\!{\it NEST, Scuola Normale Superiore, I-56126 Pisa,~Italy}

$^3$\!{\it Dipartimento di Fisica e Astronomia ``Ettore Majorana'', Universit\`a di Catania, Via S. Sofia 64, I-95123 Catania,~Italy}

$^4$\!{\it INFN, Sez.~Catania, I-95123 Catania,~Italy}

$^5$\!{\it NEST, Scuola Normale Superiore and Istituto Nanoscienze-CNR, I-56126 Pisa,~Italy}

$^6$\!{\it Department of Physics, The University of Texas at Austin, Austin, TX 78712, USA}

\bigskip

In this Supplemental Material we provide additional information on the ground-state factorization (Sec.~I), on the stiffness theorem (Sec.~II), on the TRK sum rule (Sec.~III), on the coupling of light to the EFK model degrees of freedom (Sec.~IV), on the Hartree-Fock treatment of electron-electron interactions (Sec.~V) and the resulting Bogoliubov transformation (Sec.~VI), on the $f$-sum rule (Sec.~VII), and on the phase diagram of the EFK model (Sects.~VIII and IX).
\end{center}


%

\section{Section I: Disentangling light and matter}
\label{appendix:disentagling}
In this Section we show that, in the thermodynamic $N\to \infty$ limit, it is permissible to assume a factorized ground state of the form  
\begin{equation}
\label{eq:Factorization}
\ket{\Psi}=\ket{\psi}\ket{\Phi}~.
\end{equation}
We begin by defining the electron-photon Hamiltonian
\begin{equation}
\label{eq:Hamiltonians}
\hat{\cal H}_{\rm el-ph}=\sum_{i=1}^{N} \frac{e}{m_i c} \hat{ {\bm p}}_i \cdot {\bm A}_0  (\hat{a} + \hat{a}^\dagger)+\Delta(\hat{a} + \hat{a}^\dagger)^2~,
\end{equation}
where $\Delta$ has been defined in the main text. The electron Hamiltonian $\hat{\cal H}$ and the photon Hamiltonian $\hat{\cal H}_{\rm ph}$ have been defined in the main text. Each of the Hamiltonians $\hat{\cal H}$, $\hat{\cal H}_{\rm ph}$, and $\hat{\cal H}_{\rm el-ph}$ scales {\it extensively} in $N$. While this is obvious for $\hat{\cal H}$, we note that also $\hat{\cal H}_{\rm ph}$ and $\hat{\cal H}_{\rm el-ph}$ scale with $N$ since ${\bm A}_{0}\propto 1/\sqrt{N}$ and $\hat{a}, \hat{a}^\dagger \propto \sqrt{N}$. Below, we therefore work with the operators $\hat{\cal H}/N$, $\hat{\cal H}_{\rm ph}/N$, and $\hat{\cal H}_{\rm el-ph}/N$ which are well defined in the $N\to \infty$ limit. For the sake of simplicity, we now assume that all electrons have the same mass, i.e.~$ m_{i} = m~,\forall i=1\dots N$.

In order to prove Eq.~\eqref{eq:Factorization} we will prove that, in the limit $N\to \infty$
\begin{eqnarray}
\label{eq:HamiltoniansComm}
[\frac{\hat{\cal H}}{N},\frac{\hat{\cal H}_{\rm el-ph}}{N}]\to 0
\end{eqnarray}
and
\begin{eqnarray}
\label{eq:HamiltoniansComm1}
[\frac{\hat{\cal H}_{\rm ph}}{N},\frac{\hat{\cal H}_{\rm el-ph}}{N}]\to 0~.
\end{eqnarray}
Explicitly, the left-hand side of Eq.~\eqref{eq:HamiltoniansComm} reads as following:
\begin{eqnarray}
\label{eq:HamiltoniansComm2}
[\frac{\hat{\cal H}}{N},\frac{\hat{\cal H}_{\rm el-ph}}{N}]=[\sum_{i=1}^N V(\hat{\bm r}_i)+ \frac{1}{2}\sum_{i\neq j} v(\hat{\bm r}_{i} - \hat{\bm r}_{j}),\sum_{j=1}^{N} \frac{e}{m c} \hat{ {\bm p}}_j \cdot {\bm A}_0 ] \frac{(\hat{a} + \hat{a}^\dagger)}{N^2}~.
\end{eqnarray}
Using that $[ f(\hat{\bm r}_i), \hat{ {\bm p}}_j ]=\delta_{i,j} i\hbar  \nabla_{\hat{\bm r}_i} f(\hat{\bm r}_i)$ and introducing the external force $\hat{\bm F}^{\rm ext}_i=-\nabla_{\hat{\bm r}_i}V(\hat{\bm r}_i) $ and the Coulomb force $\hat{\bm F}^{\rm C}_{i,j}=-\nabla_{\hat{\bm r}_i} v(\hat{\bm r}_{i} - \hat{\bm r}_{j})/2$ we get:
\begin{eqnarray}
\label{eq:HamiltoniansComm3}
[\frac{\hat{\cal H}}{N},\frac{\hat{\cal H}_{\rm el-ph}}{N}]=-\frac{i\hbar e (\hat{a} + \hat{a}^\dagger) {\bm A}_0}{mcN^2}\cdot \sum_{i=1}^{N} \hat{\bm F}^{\rm ext}_i~,
\end{eqnarray}
 where $\hat{\bm F}^{\rm C}_{i,j}$ dropped out of the commutator since $\sum_{i,j}\hat{\bm F}^{\rm C}_{i,j}=0$. Noticing that $(\hat{a} + \hat{a}^\dagger) {\bm A}_0$ is an {\it intensive} quantity, which does not scale with $N$, and that $\sum_{i=1}^{N} \hat{\bm F}^{\rm ext}_i \sim N$, we obtain that the commutator $[\hat{\cal H}/N,\hat{\cal H}_{\rm el-ph}/N]$ scales like $1/N$, and therefore vanishes in the thermodynamic limit.
 
Exploiting the commutator $[\hat{a},\hat{a}^\dagger]=1$, we can rewrite the left-hand side of Eq.~\eqref{eq:HamiltoniansComm1} as:
\begin{eqnarray}
\label{eq:HamiltoniansComm5}
[\frac{\hat{\cal H}_{\rm ph}}{N},\frac{\hat{\cal H}_{\rm el-ph}}{N}]=\frac{ \hbar \omega_{\rm c}}{N^2}\left\{\sum_{i=1}^{N} \frac{e}{m c} \hat{ {\bm p}}_i \cdot {\bm A}_0 (\hat{a}^\dagger-\hat{a}) +\Delta\big[ (\hat{a}^\dagger+\hat{a})(\hat{a}^\dagger-\hat{a})+(\hat{a}^\dagger-\hat{a})(\hat{a}^\dagger+\hat{a})]                 \right\}~.
\end{eqnarray}
Again, this quantity scales like $1/N$, since $ \sum_{i=1}^{N}  \hat{ {\bm p}}_i\sim N$ and $\Delta\sim 1$.

\section{Section II: On the stiffness theorem}
\label{appendix:no-gs-current}

In this Section we prove that $\braket{\psi_0|\hat{\bm j}_{\rm p}|\psi_0}\cdot {\bm A}_{0}  = 0$. We used this property to evaluate the quantity $\braket{ \psi|\hat{\cal H}|\psi}-\braket{ \psi_0|\hat{\cal H}|\psi_0}$ up to order $\bar{\beta}^2$, via the stiffness theorem.

We introduce the total dipole operator $\hat{\bm d}=-e\sum_i \hat{\bm r}_i$ and note that, because of the fundamental commutator 
$[\hat{r}_{\ell\alpha}, \hat{p}_{k\beta}]=i \hbar \delta_{\ell,k}\delta_{\alpha,\beta}$, we have
\begin{equation}\label{eq:step_1}
- i\hbar e \hat{\bm j}_{\rm p}=[\hat{\bm d} ,\hat{\cal H}]
\end{equation}
and
\begin{equation}\label{step_2}
[ \hat{j}_{{\rm p}\alpha} , \hat{d}_{\beta}]=i \hbar \sum_\ell \frac{e}{m_\ell} \delta_{\alpha,\beta}~.
\end{equation}
Using Eq.~(\ref{eq:step_1}) we immediately find for a large but finite system
\begin{eqnarray}
\braket{\psi_0|\hat{\bm j}_{\rm p}|\psi_0} \cdot {\bm A}_{0} = \frac{i}{\hbar e}\braket{\psi_0|[\hat{\bm d} ,\hat{\cal H}] |\psi_0} \cdot {\bm A}_{0}= \frac{i}{\hbar e}(E_{0} - E_{0})\braket{\psi_{0}|\hat{\bm d}|\psi_{0}} \cdot {\bm A}_{0} = 0~.
\end{eqnarray}

\section{Section III: On the TRK sum rule, i.e.~Eq.~(8) in the main text}
\label{appendix:gauge_invariance}

In this Section we prove the TRK sum rule, i.e.~Eq.~(8) in the main text.

Eq.~(\ref{eq:step_1}) implies that
\begin{eqnarray}\label{chi1}
\frac{e^2}{c^2}\sum_{n\neq0}\frac{|\langle \psi_{n}|\hat{\bm j}_{\rm p}\cdot {\bm A}_{0}|\psi_{0}\rangle|^2}{E_n-E_0}=\frac{1}{\hbar^2 c^2}\sum_{n\neq0}(E_n-E_0)  {|\langle \psi_{n}|\hat{\bm d}\cdot {\bm A}_{0}|\psi_{0}\rangle|^2}~.
\end{eqnarray}
Eq.~(\ref{step_2}) implies that we can rewrite $\Delta$ as:
\begin{eqnarray}\label{dia}
\Delta=\sum_{\ell=1}^{N} \frac{e^2 A^2_{0}}{2m_\ell c^2} =  \frac{i e}{2\hbar c^2} \langle \psi_{0}|[\hat{\bm d}\cdot {\bm A}_{0}, \hat{\bm j}_{\rm p} \cdot \bm{A}_{0}]|\psi_{0}\rangle~.
\end{eqnarray} 
We then manipulate the right-hand side of Eq.~(\ref{dia}) by inserting exact identities~\cite{Giuliani_and_VignaleS, Pines_and_NozieresS} $\openone = \sum_{n}|\psi_{n}\rangle\langle \psi_{n}|$ in the appropriate positions,
\begin{eqnarray}
\langle \psi_{0}|[\hat{\bm d}\cdot {\bm A}_{0}, \hat{\bm j}_{\rm p} \cdot \bm{A}_{0}]|\psi_{0}\rangle &=& 
\sum_{n}\langle \psi_{0}|\hat{\bm d}\cdot {\bm A}_{0}|\psi_{n}\rangle\langle \psi_{n}|\hat{\bm j}_{\rm p} \cdot \bm{A}_{0}|\psi_{0}\rangle - \sum_{n}\langle \psi_{0}|\hat{\bm j}_{\rm p} \cdot \bm{A}_{0}|\psi_{n}\rangle\langle \psi_{n}|\hat{\bm d}\cdot {\bm A}_{0}|\psi_{0}\rangle \nonumber\\
&=&\frac{i}{\hbar e}\sum_{n}\langle\psi_{0}|\hat{\bm d}\cdot {\bm A}_{0}|\psi_{n}\rangle\langle \psi_{n}|[\hat{\bm d}\cdot \bm{A}_{0}, \hat{\cal H}] |\psi_{0}\rangle -\frac{i}{\hbar e}\sum_{n}\langle\psi_{0}|[\hat{\bm d}\cdot \bm{A}_{0}, \hat{\cal H}]|\psi_{n}\rangle\langle \psi_{n}|\hat{\bm d}\cdot {\bm A}_{0} |\psi_{0}\rangle\nonumber\\
&=&-\frac{2i}{\hbar e}\sum_{n}(E_{n}-E_{0})
|\langle \psi_{n}|\hat{\bm d}\cdot \bm{A}_{0}|\psi_{0}\rangle|^2 =-\frac{2i}{\hbar e}
\sum_{n\neq 0}(E_{n}-E_{0})
|\langle \psi_{n}|\hat{\bm d}\cdot \bm{A}_{0}|\psi_{0}\rangle|^2~.
\end{eqnarray} 
Using the previous result inside Eq.~(\ref{dia}), we find
\begin{eqnarray}\label{dia_final}
\Delta=\frac{1}{\hbar^2 c^2}\sum_{n\neq 0}(E_{n}-E_{0})
|\langle \psi_{n}|\hat{\bm d}\cdot \bm{A}_{0}|\psi_{0}\rangle|^2~.
\end{eqnarray} 
Comparing Eq.~(\ref{dia_final}) with Eq.~(\ref{chi1}) we reach the desired result, i.e.~Eq.~(8) of the main text.

We now present a more physical, alternative proof.  We first remind the reader that the physical  current operator corresponding to the Hamiltonian $\hat{\cal H}_{{\bm A}_{0}}$, Eq.~(2) in the main text, is
\begin{equation}
\hat{\bm J}_{\rm phys} = \frac{c}{e}\frac{\delta \hat{\cal H}_{{\bm A}_{0}}}{\delta {\bm A}_0}=\hat{\bm j}_{\rm p}+ \sum_{i=1}^{N} \frac{e}{m_{i}c}{\bm A}_{0}~.
\end{equation}
We now observe that the electron system cannot respond to ${\bm A}_{0}$, since the latter is uniform and time-independent. (A current cannot flow along ${\bm u}$ in response to ${\bm A}_{0}$.) This property, i.e.~gauge invariance, implies that the physical current-current response function in response to ${\bm A}_{0}$ must vanish~\cite{Giuliani_and_VignaleS,Pines_and_NozieresS}, i.e.
\begin{equation}\label{eq:gauge_invariance}
 0 = -\frac{2}{L^d}\sum_{n\neq0}\frac{|\langle \psi_{n}|\hat{\bm j}_{\rm p}\cdot {\bm u}|\psi_{0}\rangle|^2}{E_{n}-E_0} + \frac{1}{L^d}\sum_{i=1}^{N} \frac{1}{m_{i}}~,
\end{equation}
where the first (second) term on the right-hand side is the paramagnetic (diamagnetic) contribution and $L^d$ is the electron system volume. Eq.~(\ref{eq:gauge_invariance}) can be written as
\begin{equation}
2\sum_{n\neq0}\frac{|\langle \psi_{n}|\hat{\bm j}_{\rm p}\cdot {\bm u}|\psi_{0}\rangle|^2}{E_{n}-E_0} = \sum_{i=1}^{N} \frac{1}{m_i}~,
\end{equation}
which is easily seen to be equivalent to Eq.~(8). 
In other words,  Eq.~(8) simply expresses the fact that paramagnetic and diamagnetic contributions to the physical current-current response function cancel out in the uniform and static limit~\cite{Giuliani_and_VignaleS, Pines_and_NozieresS}.

\section{Section IV: Coupling the EFK model to cavity photons}
Consider spinless electrons hopping in a one-dimensional inversion-symmetric crystal with $N$ sites, one atom per site, and two atomic orbitals of opposite parity (${\rm s}$ and ${\rm p}$), in a tight-binding scheme. The second-quantized single-particle Hamiltonian in the site representation reads as following:
\begin{eqnarray}\label{eq:1D_semiconductor}
\hat{\cal H}_{0}&=&\sum_{j=1}^{N} \sum_{\alpha = {\rm s}, {\rm p}}E_{\alpha} \hat{c}^\dagger_{j, \alpha}\hat{c}_{j,  \alpha} -t_{\rm s} \sum_{j=1}^{N} (\hat{c}^\dagger_{j+1, {\rm s}}\hat{c}_{j,  {\rm s}}+\hat{c}^\dagger_{j,  {\rm s}}\hat{c}_{j+1, {\rm s}})+t_{\rm p} \sum_{j=1}^{N} (\hat{c}^\dagger_{j+1, {\rm p}}\hat{c}_{j,  {\rm p }}+\hat{c}^\dagger_{j,  {\rm p }}\hat{c}_{j+1, {\rm p}}) \nonumber \\ 
&-&\tilde{t}\sum_{j=1}^{N} (\hat{c}^\dagger_{j+1, {\rm s}}\hat{c}_{j,  {\rm p}}  +\hat{c}^\dagger_{j,  {\rm p}}\hat{c}_{j+1, {\rm s}})+\tilde{t} \sum_{j=1}^{N} (\hat{c}^\dagger_{j, {\rm s}}\hat{c}_{j+1,  {\rm p }}+\hat{c}^\dagger_{j+1,  {\rm p }}\hat{c}_{j, {\rm s}}) \equiv \sum_{j=1}^{N} \sum_{\alpha = {\rm s}, {\rm p}}E_{\alpha} \hat{c}^\dagger_{j, \alpha}\hat{c}_{j,  \alpha} + \hat{\cal T}~,
\end{eqnarray}
where $t_{\rm s}$,  $t_{\rm p}$, $\tilde{t}\in \mathbb{R}$ are for the moment completely arbitrary, we assumed periodic boundary conditions ($ \hat{c}_{N+1, \alpha} =\hat{c}_{1,\alpha}$), and defined the kinetic operator $\hat{\cal T}=t_{\rm p} \sum_{j=1}^{N} (\hat{c}^\dagger_{j+1, {\rm p}}\hat{c}_{j,  {\rm p }}+\hat{c}^\dagger_{j,  {\rm p }}\hat{c}_{j+1, {\rm p}}) -\tilde{t}\sum_{j=1}^{N} (\hat{c}^\dagger_{j+1, {\rm s}}\hat{c}_{j,  {\rm p}}  +\hat{c}^\dagger_{j,  {\rm p}}\hat{c}_{j+1, {\rm s}})+\tilde{t} \sum_{j=1}^{N} (\hat{c}^\dagger_{j, {\rm s}}\hat{c}_{j+1,  {\rm p }}+\hat{c}^\dagger_{j+1,  {\rm p }}\hat{c}_{j, {\rm s}})$. We now add repulsive on-site electron-electron interactions in the site representation:
\begin{equation}\label{eq:Hubbard-U}
\hat{\cal H}_{\rm ee} = U \sum_{j=1}^{N} \hat{n}_{j,s}\hat{n}_{j,p}~,
\end{equation}
where $U>0$ and $\hat{n}_{j,\alpha} \equiv \hat{c}^\dagger_{j,\alpha} \hat{c}_{j,\alpha}$ is the orbitally-resolved local density operator.

The full Hamiltonian of our 1D EFK model in the absence of cavity photons is
\begin{equation}\label{eq:complete}
\hat{\cal H} = \hat{\cal H}_{0} + \hat{\cal H}_{\rm ee}~.
\end{equation}

We now couple the matter Hamiltonian $\hat{\cal H}$ in Eq.~(\ref{eq:complete}) to light by employing a uniform linearly-polarized vector potential ${\bm A}(t) = A(t)\hat{\bm u}$ where ${\bm u} = \pm \hat{\bm x}$ in the ring geometry above with periodic boundary conditions. This is accomplished, as usual~\cite{kohn_pr_1964S,shastry_prl_1990S,millis_prb_1990S,fye_prb_1991S}, by means of the Peierls factor:
\begin{eqnarray}
\hat{\cal H}_{A(t)}&=&\sum_{j=1}^{N} \sum_{\alpha = {\rm s}, {\rm p}}E_{\alpha} \hat{c}^\dagger_{j, \alpha}\hat{c}_{j,  \alpha} -t_{\rm s} \sum_{j=1}^{N} \Big(e^{-i e a A(t)/(\hbar c)}\hat{c}^\dagger_{j+1, {\rm s}}\hat{c}_{j,  {\rm s}}+e^{+i e a A(t)/(\hbar c)}\hat{c}^\dagger_{j,  {\rm s}}\hat{c}_{j+1, {\rm s}} \Big)\nonumber\\
&+&t_{\rm p} \sum_{j=1}^{N} \Big(e^{-i e a A(t)/(\hbar c)}\hat{c}^\dagger_{j+1, {\rm p}}\hat{c}_{j,  {\rm p }}+e^{+i e a A(t)/(\hbar c)}\hat{c}^\dagger_{j,  {\rm p }}\hat{c}_{j+1, {\rm p}} \Big) \nonumber \\ 
&-&\tilde{t}\sum_{j=1}^{N} \Big(e^{-i e a A(t)/(\hbar c)}\hat{c}^\dagger_{j+1, {\rm s}}\hat{c}_{j,  {\rm p}}  + e^{+i e a A(t)/(\hbar c)}\hat{c}^\dagger_{j,  {\rm p}}\hat{c}_{j+1, {\rm s}}\Big)\nonumber\\
&+&\tilde{t} \sum_{j=1}^{N} \Big(e^{+i e a A(t)/(\hbar c)}\hat{c}^\dagger_{j, {\rm s}}\hat{c}_{j+1,  {\rm p }}+e^{-i e a A(t)/(\hbar c)}\hat{c}^\dagger_{j+1,  {\rm p }}\hat{c}_{j, {\rm s}}\Big)\nonumber\\
&+&\hat{\cal H}_{\rm ee}~,
\end{eqnarray}
where $a$ is the lattice constant.

We expand $\hat{\cal H}_{A(t)}$ in powers of $A(t)$ for small $A(t)$, retaining terms of ${\cal O}(A^2(t))$. We find:
\begin{equation}
\hat{\cal H}_{A(t)}=\hat{\cal H}_{0}+\hat{\cal H}_{\rm ee}+ \frac{e}{c}A(t)\hat{j}_{\rm p}-\frac{1}{2}\frac{e^2 a^2 A^2(t)}{\hbar^2c^2} \hat{\cal T}~,
\end{equation}
where
\begin{eqnarray}
\hat{j}_{\rm p} \equiv \frac{c}{e}\left.\frac{\delta \hat{\cal H}_{A(t)}}{\delta A(t)}\right|_{A(t)=0} &=& \frac{it_{\rm s} a}{\hbar}\sum_{j=1}^{N} (\hat{c}^\dagger_{j+1, {\rm s}}\hat{c}_{j,  {\rm s}}-\hat{c}^\dagger_{j,  {\rm s}}\hat{c}_{j+1, {\rm s}}) -\frac{it_{\rm p} a}{\hbar}\sum_{j=1}^{N} (\hat{c}^\dagger_{j+1, {\rm p}}\hat{c}_{j,  {\rm p }}-\hat{c}^\dagger_{j,  {\rm p }}\hat{c}_{j+1, {\rm p}} )\nonumber\\
&+&\frac{i\tilde{t} a}{\hbar}\sum_{j=1}^{N} (\hat{c}^\dagger_{j+1, {\rm s}}\hat{c}_{j,  {\rm p}}  - \hat{c}^\dagger_{j,  {\rm p}}\hat{c}_{j+1, {\rm s}})+\frac{i\tilde{t} a}{\hbar} \sum_{j=1}^{N} (\hat{c}^\dagger_{j, {\rm s}}\hat{c}_{j+1,  {\rm p }}- \hat{c}^\dagger_{j+1,  {\rm p }}\hat{c}_{j, {\rm s}})
\end{eqnarray}
is the paramagnetic (number) current operator and $\hat{\cal T}$ is the kinetic operator in Eq.~(\ref{eq:1D_semiconductor}). The physical (number) current operator is therefore
\begin{equation}\label{eq:physical_current_lattice_model}
\hat{J}_{\rm phys} \equiv \frac{c}{e}\frac{\delta \hat{\cal H}_{A(t)}}{\delta A(t)} = \hat{j}_{\rm p} - \frac{e}{c} A(t)\frac{a^2}{\hbar^2}\hat{\cal T}~,
\end{equation}
with contains paramagnetic and diamagnetic terms.

We finally quantize the e.m. field by writing $A(t) \to A_{0}(\hat{a} +\hat{a}^\dagger)$, where $A_{0}$ has been defined in the main text, and we give dynamics to the field by means of the photon Hamiltonian $\hat{\cal H}_{\rm ph} = \hbar\omega_{\rm c} \hat{a}^\dagger \hat{a}$. The full Hamiltonian, which includes light-matter interactions, is therefore given by:
\begin{equation}\label{eq:quadratic_Hamiltonian_final_site_representation}
\hat{\cal H}_{{\bm A}_{0}}=\hat{\cal H}_{0}+\hat{\cal H}_{\rm ee} + \hat{\cal H}_{\rm ph}+ \frac{e}{c}A_{0} \hat{j}_{\rm p}(\hat{a} + \hat{a}^\dagger)-\frac{1}{2}\frac{e^2 a^2}{\hbar^2c^2}A^2_{0} \hat{\cal T} (\hat{a} + \hat{a}^\dagger)^2~.
\end{equation}
The fourth and fifth terms in the right-hand side of Eq.~(\ref{eq:quadratic_Hamiltonian_final_site_representation}) are called paramagnetic and diamagnet contributions.

Eq.~(\ref{eq:quadratic_Hamiltonian_final_site_representation}) is written in the site representation. In the main text, however, the quantities $\hat{\cal H}_{0}$, $\hat{j}_{\rm p}$, and $\hat{\cal T}$ have been given in momentum space.
The link between momentum-space and site representations is offered by
\begin{equation}
\hat{c}^\dagger_{j,\alpha} = \frac{1}{\sqrt{N}}\sum_{k \in {\rm BZ}} \hat{c}^\dagger_{k,\alpha} e^{-i k j a}~,
\end{equation}
where the sum is carried over the 1D Brillouin zone (BZ). In the thermodynamic $N\to \infty$ limit we can replace
\begin{equation}\label{eq:t-limit}
\frac{1}{N}\sum_{k\in{\rm BZ}}\to a \int_{-\pi/a}^{+\pi/a}\frac{dk}{2\pi}~. 
\end{equation}
We find
\begin{equation}\label{eq:1D_semiconductor_FT}
\hat{\cal H}_{0}
=\sum_{k\in {\rm BZ}}
 \begin{pmatrix}
\hat{c}^\dagger_{k, {\rm s}} & \hat{c}^\dagger_{k, {\rm p}}
\end{pmatrix}
 \begin{pmatrix}
E_{\rm s} -2 t_{\rm s} \cos(ka) & 2 i \tilde{t} \sin(ka)\\
-2 i \tilde{t} \sin(ka) &  E_{\rm p} + 2 t_{\rm p} \cos(ka)
\end{pmatrix}
 \begin{pmatrix}
\hat{c}_{k, {\rm s}} \\ \hat{c}_{k, {\rm p}}
\end{pmatrix}
\equiv \sum_{k\in {\rm BZ}}\sum_{\alpha,\beta={\rm s}, {\rm p}}\hat{c}^\dagger_{k,\alpha}H_{\alpha\beta}(k)\hat{c}_{k,\beta}~,
\end{equation}
\begin{equation}\label{eq:paracurrent_FT}
\hat{j}_{\rm p} 
=\frac{2a}{\hbar}\sum_{k\in {\rm BZ}}
 \begin{pmatrix}
\hat{c}^\dagger_{k, {\rm s}} & \hat{c}^\dagger_{k, {\rm p}}
\end{pmatrix}
 \begin{pmatrix}
t_{\rm s}\sin(ka) &i\tilde{t}\cos(ka)\\
-i\tilde{t}\cos(ka) &-t_{\rm p}\sin(ka)
\end{pmatrix}
 \begin{pmatrix}
\hat{c}_{k, {\rm s}} \\ \hat{c}_{k, {\rm p}}
\end{pmatrix}\equiv \sum_{k\in {\rm BZ}}\sum_{\alpha,\beta={\rm s}, {\rm p}}\hat{c}^\dagger_{k,\alpha}j_{\alpha\beta}(k)\hat{c}_{k,\beta}~,
\end{equation}
and
\begin{equation}\label{eq:kineticoperator_FT}
\hat{\cal T} = 2 \sum_{k\in {\rm BZ}}
 \begin{pmatrix}
\hat{c}^\dagger_{k, {\rm s}} & \hat{c}^\dagger_{k, {\rm p}}
\end{pmatrix}
 \begin{pmatrix}
-t_{\rm s}\cos(ka) &i\tilde{t}\sin(ka)\\
-i\tilde{t}\sin(ka) &+t_{\rm p}\cos(ka)
\end{pmatrix}
 \begin{pmatrix}
\hat{c}_{k, {\rm s}} \\ \hat{c}_{k, {\rm p}}
\end{pmatrix}
\equiv \sum_{k\in {\rm BZ}}\sum_{\alpha,\beta={\rm s}, {\rm p}}\hat{c}^\dagger_{k,\alpha}{\cal T}_{\alpha\beta}(k)\hat{c}_{k,\beta}~.
\end{equation}
It is easy to check that
\begin{equation}\label{eq:key_relation_1}
j_{\alpha\beta}(k) = \frac{1}{\hbar}\frac{\partial H_{\alpha\beta}(k)}{\partial k}
\end{equation}
and
\begin{equation}\label{eq:key_relation_2}
{\cal T}_{\alpha\beta}(k) = -\frac{1}{a^2}\frac{\partial^2 H_{\alpha\beta}(k)}{\partial k^2}~.
\end{equation}
Equations~(\ref{eq:key_relation_1})-(\ref{eq:key_relation_2}) heavily constrain the paramagnetic and diamagnetic terms of the full Hamiltonian $\hat{\cal H}_{{\bm A}_{0}}$, which rule light-matter interactions. In other words, one cannot simply couple light to matter with arbitrary operators $\hat{j}_{\rm p}$ and $\hat{\cal T}$.  Instead the form 
of these operators is specified by $\hat{\cal H}_0$ and must be constructed with perfect consistency.

\section{Section V: Hartree-Fock treatment of electron-electron interactions}
\label{sect:meanfieldtheory}

We treat the electron-electron interaction term in Eq.~(11) of the main text---or, equivalently, Eq.~(\ref{eq:Hubbard-U})---within Hartree-Fock (HF) mean-field theory (see, e.g., Chapter 2 of Ref.~\cite{Giuliani_and_VignaleS}). We replace $\hat{\cal H}_{\rm ee}$ with
\begin{align}\label{eq:MFdecoupling}
\hat{\cal H}^{({\rm HF})}_{\rm ee} &\equiv U \sum_{j=1}^{N} 
[\hat{c}^\dag_{j, {\rm s}} \hat{c}_{j, {\rm s}}  
\langle  \hat{c}^\dag_{j, {\rm p}} \hat{c}_{j, {\rm p}} \rangle +
\hat{c}^\dag_{j, {\rm p}} \hat{c}_{j, {\rm p}}  \langle  
\hat{c}^\dag_{j, {\rm s}} \hat{c}_{j,{\rm s}} \rangle 
- \hat{c}^\dag_{j, {\rm s}} \hat{c}_{j, {\rm p}}  \langle \hat{c}^\dag_{j, {\rm p}} \hat{c}_{j, {\rm s}}  \rangle 
- \hat{c}^\dag_{j, {\rm p}} \hat{c}_{j, {\rm s}}  \langle \hat{c}^\dag_{j, {\rm s}} \hat{c}_{j, {\rm p}}  \rangle ]\nonumber \\
&-U \sum_{i} [ \langle  \hat{c}^\dag_{j, {\rm s}} \hat{c}_{j, {\rm s}} \rangle \langle  \hat{c}^\dag_{j, {\rm p}} \hat{c}_{j, {\rm p}} \rangle - \langle \hat{c}^\dag_{j, {\rm s}} \hat{c}_{j, {\rm p}}  \rangle \langle \hat{c}^\dag_{j, {\rm p}} \hat{c}_{j, {\rm s}}  \rangle ]~.
\end{align}
Each of the mean fields above can be written as
\begin{equation}
\langle  \hat{c}^\dag_{j,\alpha} \hat{c}_{j,\beta} \rangle = \frac{1}{N}\sum_{k, k'\in {\rm BZ}} e^{-i(k-k^\prime)ja}  \langle  \hat{c}^\dag_{k,\alpha} \hat{c}_{k^\prime,\beta} \rangle~.
\end{equation}
We assume that $\langle  \hat{c}^\dag_{j,\alpha} \hat{c}_{j,\beta} \rangle$ is independent of the site index $j$ (translational invariance), i.e.~we take $\langle  \hat{c}^\dag_{k,\alpha} \hat{c}_{k^\prime,\beta} \rangle=\delta_{k,k'}\langle  \hat{c}^\dag_{k,\alpha} \hat{c}_{k,\beta} \rangle$. 

We are therefore naturally led to introduce the following quantities:
\begin{equation}
{\cal M} \equiv \langle \hat{c}^\dag_{j, {\rm p}} \hat{c}_{j, {\rm p}} \rangle - \langle  \hat{c}^\dag_{j,{\rm s}} \hat{c}_{j,{\rm s}} \rangle =
\frac{1}{N} \sum_{k\in {\rm BZ}} ( \langle  \hat{c}^\dag_{k,{\rm p}} \hat{c}_{k,{\rm p}} \rangle - \langle  \hat{c}^\dag_{k,{\rm s}} \hat{c}_{k,{\rm s}} \rangle)~,
\end{equation}
\begin{equation}
{\cal I} \equiv \langle  \hat{c}^\dag_{j,{\rm p}} \hat{c}_{j,{\rm s}} \rangle=
\frac{1}{N} \sum_{k \in {\rm BZ}} \langle  \hat{c}^\dag_{k,{\rm p}} \hat{c}_{k,{\rm s}} \rangle~,
\end{equation}
and
\begin{equation}
n_0 \equiv \langle  \hat{c}^\dag_{j,{\rm p}} \hat{c}_{j,{\rm p}} \rangle + \langle  \hat{c}^\dag_{j,{\rm s}} \hat{c}_{j,{\rm s}} \rangle = 
\frac{1}{N} \sum_{k \in {\rm BZ}}( \langle  \hat{c}^\dag_{k,{\rm p}} \hat{c}_{k,{\rm p}} \rangle+ \langle  \hat{c}^\dag_{k,{\rm s}} \hat{c}_{k,{\rm s}} \rangle)~.
\end{equation}
Under the assumption of homogeneity, we can rewrite the HF interaction term (\ref{eq:MFdecoupling}) as
\begin{align}\label{eq:VMF*S}
\hat{\cal H}^{({\rm HF})}_{\rm ee} &= -U \sum_{k\in {\rm BZ}} 
\left[
\frac{\cal M}{2 }(\hat{c}^\dag_{k,{\rm p}} \hat{c}_{k,{\rm p}} - \hat{c}^\dag_{k,{\rm s}} \hat{c}_{k,{\rm s}})+
{\cal I} \hat{c}^\dag_{k,{\rm s}} \hat{c}_{k,{\rm p}}  
+ {\cal I}^\ast  \hat{c}^\dag_{k,{\rm p}} \hat{c}_{k,{\rm s}} 
\right]  \nonumber\\
& +U \sum_{k \in {\rm BZ}} 
\frac{n_0}{2}(\hat{c}^\dag_{k,{\rm p}} \hat{c}_{k,{\rm p}}+\hat{c}^\dag_{k,{\rm s}} \hat{c}_{k,{\rm s}}) 
+U N \left( \frac{{\cal M}^2-n^2_{0}}{4} + |{\cal I}|^2 \right)~.
\end{align}
The term proportional to $n_{0}/2$ acts as a renormalization of the chemical potential $\mu$ in the grand-canonical Hamiltonian $\hat{\cal K} = \hat{\cal H}_{{\bm A}_{0}} - \mu \hat{\cal N}$, where $\hat{\cal N} = \sum_{k\in {\rm BZ}}\sum_{\alpha={\rm s}, {\rm p}}\hat{c}^\dagger_{k,\alpha}\hat{c}_{k,\alpha}$ is the total electron number operator. In this work we study only the phase diagram at half filling. We therefore have $n_{0}=1~\forall j =1 \dots N$ in all phases and can discard such term. The last term in Eq.~(\ref{eq:VMF*}) instead must be retained (after discarding $n_{0}$) since it takes different values in different phases. (It is a trivial constant: it therefore only matters when one compares total energies of different phases.)

The HF mean-field Hamiltonian (\ref{eq:VMF*S}) can be written in a $2\times 2$ fashion:
\begin{equation}\label{eq:eeMF}
\hat{\cal H}^{({\rm HF})}_{\rm ee}=U\sum_{k\in {\rm BZ}}
 \begin{pmatrix}
\hat{c}^\dagger_{k, {\rm s}} & \hat{c}^\dagger_{k, {\rm p}}
\end{pmatrix}
 \begin{pmatrix}
{\cal M}/2 & -{\cal I}\\
-{\cal I}^\ast & -{\cal M}/2
\end{pmatrix}
 \begin{pmatrix}
\hat{c}_{k, {\rm s}} \\ \hat{c}_{k, {\rm p}}
\end{pmatrix}
+U N \left( \frac{{\cal M}^2}{4} + |{\cal I}|^2 \right)~.
\end{equation}
\section{Section VI: Details on the Bogoliubov transformation}

In this Section we give all technical details relevant to the diagonalization of the problem posed by Eq.~(9) in the main text, with $\hat{\cal H}_{\rm ee}$ replaced by its HF mean-field expression (\ref{eq:eeMF}):
\begin{equation}
\hat{\cal H}^{({\rm HF})}_{{\bm A}_{0}}\equiv\hat{\cal H}_{0}+\hat{\cal H}^{({\rm HF})}_{\rm ee} + \hat{\cal H}_{\rm ph}+ \frac{g_{0}}{\sqrt{N}}\frac{\hbar}{a} \hat{j}_{\rm p}(\hat{a} + \hat{a}^\dagger)-\frac{g^2_{0}}{2N}\hat{\cal T} (\hat{a} + \hat{a}^\dagger)^2~.
\end{equation}
We seek ground-state wave functions of the unentangled form $\ket{\Psi}=\ket{\psi}\ket{\Phi}$, where $\ket{\psi}$ is the wave function for the matter degrees of freedom and $\ket{\Phi}$ is the analog for the e.m. field. 

In order to reduce the number of free parameters in the problem, we enforce particle-hole symmetry by setting $E_{\rm s}=-E_{\rm g}/2=-E_{\rm p}$ and $t_{\rm s}=t_{\rm p}=t$.

An effective mean-field Hamiltonian for matter degrees of freedom can be obtained by taking the expectation value of $\hat{\cal H}^{({\rm HF})}_{{\bm A}_{0}}$ over the light state $\ket{\Phi}$, i.e.
\begin{eqnarray}\label{eq:eff-H-MF}
\hat{\cal H}_{\rm eff-matter} \equiv \langle\Phi|\hat{\cal H}^{({\rm HF})}_{{\bm A}_{0}}|\Phi\rangle
&=&\sum_{k\in {\rm BZ}}
 \begin{pmatrix}
\hat{c}^\dagger_{k, {\rm s}} & \hat{c}^\dagger_{k, {\rm p}}
\end{pmatrix}
{\cal H}(k)
\begin{pmatrix}
\hat{c}_{k, {\rm s}} \\ \hat{c}_{k, {\rm p}}
\end{pmatrix}
+\hbar\omega_{\rm c} \braket{\Phi|   a^\dagger a|\Phi}+UN \left( \frac{{\cal M}^2}{4} + |{\cal I}|^2 \right)~.
\end{eqnarray}
${\cal H}(k)$ can be conveniently written in terms of ordinary $2\times 2$ Pauli matrices $\{\sigma_{i}, i=1,2,3\}$, i.e.~${\cal H}(k)=\sum_i h_i(k)\sigma_i$ with
\begin{equation}\label{Hk_1}
h_1(k)=-U{\rm Re}({\cal I})~,
\end{equation}
\begin{equation}\label{Hk_2}
h_2(k)=- 2 \tilde{t} \sin(ka) \left(1-\frac{g^2_{0}}{2}{\cal A}_{2}\right) -2\tilde{t}g_{0}\cos(ka){\cal A}_{1}+U{\rm Im}({\cal I})~,
\end{equation}
and
\begin{equation}\label{Hk_3}
h_3(k)=- \frac{E_{\rm g}}{2} -2t\cos(ka)\left(1-\frac{g^2_{0}}{2}{\cal A}_2\right) +2t g_{0}\sin(ka){\cal A}_{1}+U\frac{\cal M}{2}~.
\end{equation}
In Eqs.~(\ref{Hk_1})-(\ref{Hk_3}) we have introduced
\begin{equation}\label{eq:A1}
{\cal A}_1\equiv \frac{1}{\sqrt{N}}\braket{\Phi|\hat{a} + \hat{a}^\dagger|\Phi}
\end{equation}
and
\begin{equation}\label{eq:A2}
{\cal A}_2\equiv \frac{1}{N}\braket{\Phi|\left(\hat{a}+\hat{a}^\dagger\right)^2|\Phi}~.
\end{equation}

The Hamiltonian (\ref{eq:eff-H-MF}) can be diagonalized by introducing the following Bogoliubov transformation:
\begin{eqnarray}
\hat{\gamma}^\dagger_{k,-}=u_{k} \hat{c}^\dagger_{k,{\rm s}}+v_{k} \hat{c}^\dagger_{k,{\rm p}}~,\\
\hat{\gamma}^\dagger_{k,+}=v^*_{k} \hat{c}^\dagger_{k,{\rm s}}-u^*_{k} \hat{c}^\dagger_{k,{\rm p}}~,
\end{eqnarray}
where $u_k =\cos(\theta_{k}/2)$ and $v_k=\sin(\theta_{k}/2)e^{i\phi_{k}}$ with
\begin{eqnarray}
\cos(\theta_{k})&=&-\frac{h_{3}(k)}{\epsilon(k)}~,\\
\sin(\theta_{k})&=&-\frac{\sqrt{h^2_{1}(k)+h^2_{2}(k)}}{\epsilon(k)}~,\\
e^{i\phi_k}&=&\frac{{h_{1}(k)+i h_{2}(k)}} {\sqrt{h^2_{1}(k)+h^2_{2}(k)}}~,\\
\epsilon(k)&=&\sqrt{h^2_{1}(k)+h^2_{2}(k)+h^2_{3}(k)}\label{eq:spettro}~.
\end{eqnarray}
Note that $u_{k}$ and $v_{k}$ are functions of ${\cal I}$, ${\cal M}$, ${\cal A}_1$, and ${\cal A}_2$, i.e.~$u_k =u_k({\cal I}, {\cal M}, {\cal A}_1, {\cal A}_2)$ and $v_k =v_k({\cal I}, {\cal M}, {\cal A}_1, {\cal A}_2)$. We find
\begin{eqnarray}\label{eq:Bogoliubov_Hamiltonian}
\hat{\cal H}_{\rm eff-matter} =\sum_{k\in {\rm BZ}} \sum_{\xi=\pm} \xi \epsilon(k)\hat{\gamma}^\dagger_{k,\xi}\hat{\gamma}_{k,\xi}+ {\cal C}~,
\end{eqnarray}
where
\begin{equation}\label{eq:cost}
{\cal C}\equiv  U N \left( \frac{{\cal M}^2}{4} + |{\cal I}|^2 \right) + \hbar\omega_{\rm c}\braket{\Phi|   a^\dagger a|\Phi}~.
\end{equation}
The ground state of (\ref{eq:Bogoliubov_Hamiltonian}) is $\ket{\psi}=\prod_{k\in {\rm BZ}} \hat{\gamma}^\dagger_{k,-}\ket{\rm vac}$, where $\ket{\rm vac}$ is the state with no electrons. Mimicking the BCS theory, we find that
\begin{equation}\label{eq:SBCS}
\ket{\psi}=\prod_{k\in {\rm BZ}} \big [  u_{k} +v_{k} \hat{c}^\dagger_{k,{\rm p}} \hat{c}_{k,{\rm s}} \big]  \ket{\emptyset}~,
\end{equation}
where $\ket{\emptyset}=\prod_{k\in {\rm BZ}} \hat{c}^\dagger_{k,{\rm s}}\ket{\rm vac}$.
The following quantities $\braket{\psi |\hat{c}^\dagger_{k,{\rm s}}\hat{c}_{k,{\rm s}}|\psi}= |u_{k}|^2$, $\braket{\psi| \hat{c}^\dagger_{k,{\rm p}}\hat{c}_{k,{\rm p}}|\psi}= |v_{k}|^2$, and $\braket{\psi | \hat{c}^\dagger_{k,{\rm p}}\hat{c}_{k,{\rm s}}|\psi}= v^*_{k}u_{k}$ are useful to write the order parameters in terms of $u_{k}$ and $v_{k}$. We find
\begin{equation}\label{gapeq-I}
{\cal I} =  \frac{1}{N} \sum_{k \in {\rm BZ}} \braket{\psi|\hat{c}^\dagger_{k,{\rm p}}\hat{c}_{k,{\rm s}}|\psi } = \frac{1}{N} \sum_{k\in {\rm BZ}} v^*_{k}u_{k}
\end{equation}
and
\begin{equation}\label{gapeq-M}
{\cal M} = \frac{1}{N} \sum_{k\in {\rm BZ}} (\braket{\psi|\hat{c}^\dagger_{k,{\rm p}}\hat{c}_{k,{\rm p}}|\psi} -   \braket{\psi|\hat{c}^\dagger_{k,{\rm s}}\hat{c}_{k,{\rm s}}|\psi})=\frac{1}{N} \sum_{k\in {\rm BZ}}  \big(|v_{k}|^2-|u_{k}|^2\big)~.
\end{equation}
We also write the expectation values of $\hat{j}_{\rm p}$ and $\hat{\cal T}$ over the HF state $\ket{\psi}$ in terms of $u_{k}$ and $v_{k}$:
\begin{equation}\label{gapeq2-J}
{\cal J} \equiv \frac{\hbar}{a N}\braket{\psi|\hat{j}_{\rm p}|\psi} =\frac{2}{N}\sum_{k\in {\rm BZ}}\left[- t\sin(ka)(|v_{k}|^2-|u_{k}|^2)- 2\tilde{t}\cos(ka){\rm Im}(u^\ast_{k}v_{k}) \right]
\end{equation}
and
\begin{equation}\label{gapeq2-T}
{\cal T} \equiv \braket{\psi|\hat{\cal T}|\psi}=\frac{2}{N}  \sum_{k\in {\rm BZ}}\left[ t\cos(ka)   \big(|v_{k}|^2-|u_{k}|^2\big)-  2\tilde{t}\sin(ka) {\rm Im}(u^*_{k}v_{k}) \right]~.
\end{equation}
Note that both ${\cal J}$ and ${\cal T}$ have units of energy and are finite in the thermodynamic $N\to \infty$ limit.

Following exactly the same steps described in the proof of the no-go theorem in the main text and defining
\begin{equation}
\Delta=-\frac{g^2_{0}}{2}{\cal T}
\end{equation}
and
\begin{equation}
\lambda=\sqrt{1+\frac{4\Delta}{\hbar \omega_{\rm c}}}~,
\end{equation}
we find that $\ket{\Phi}$ must be a coherent state $\ket{\bar{\beta}}$, i.e.~$\hat{b}\ket{\bar{\beta}}=\bar{\beta}\ket{\bar{\beta}}$, with
\begin{eqnarray}
\bar{\beta}=-\frac{g_{0}}{\lambda^{3/2}}\frac{\cal J}{\hbar\omega_{\rm c}}\sqrt{N}~, 
\end{eqnarray}
to be compared with Eq.~(4) in the main text. As in the case of the proof of the no-go theorem, $\hat{b} = \cosh(x) \hat{a} + \sinh(x) \hat{a}^\dag$, with $\cosh(x) = (\lambda+1)/(2\sqrt{\lambda})$ and 
$\sinh(x) = (\lambda-1)/(2\sqrt{\lambda})$. The inverse transformation reads as following: $\hat{a}=\cosh(x)\hat{b}-\sinh(x)\hat{b}^\dagger$. Note that ${\cal J}$ depends on $\bar{\beta}$ and therefore the previous equation defines $\bar{\beta}$ only implicitly.

Since we have found the ground state $\ket{\Phi}$ (i.e.~a coherent state $\ket{\bar{\beta}}$ of the $\hat{b}$ operator), we can evaluate Eqs.~(\ref{eq:A1})-(\ref{eq:A2}):
\begin{equation}\label{min-A1}
{\cal A}_{1} =\frac{1}{\sqrt{N\lambda}}\langle\bar{\beta}|\hat{b} + \hat{b}^\dagger|\bar{\beta}\rangle = \frac{2\bar{\beta}}{\sqrt{N\lambda}}=-\frac{2g_{0}}{\lambda^2}\frac{\cal J}{\hbar\omega_{\rm c}}
\end{equation}
and
\begin{equation}\label{min-A2}
{\cal A}_{2} =\frac{1}{N\lambda}\braket{\bar{\beta}|\big(\hat{b} +\hat{b}^\dagger\big)^2|\bar{\beta}}=\frac{4\bar{\beta}^2+1}{N\lambda}={\cal A}^2_{1}+\frac{1}{\lambda N}~.
\end{equation}
To derive Eq.~(\ref{min-A2}) we have used that $\big(\hat{b} +\hat{b}^\dagger\big)^2 = \hat{b}^2 +\hat{b}^{\dagger2}+2\hat{b}^{\dagger}\hat{b}+1$. Using Eqs.~(\ref{eq:Bogoliubov_Hamiltonian})-(\ref{eq:cost}) and using that $\hat{a}^\dagger\hat{a} = (\lambda^2+1)\hat{b}^\dagger\hat{b}/(2\lambda) -(\lambda^2-1)(\hat{b}^2+\hat{b}^{\dagger2})/(4\lambda)+(\lambda-1)^2/(4\lambda)$, we can also write the ground-state energy per particle as
\begin{equation}
\epsilon_{\rm GS} = \frac{E_{\rm GS}}{N} = - \frac{1}{N}\sum_{k\in {\rm BZ}} \epsilon(k) +\frac{\hbar\omega_{\rm c}[4\bar{\beta}^2+(\lambda-1)^2]}{4\lambda N}+U \left( \frac{{\cal M}^2}{4} + |{\cal I}|^2 \right)~.
\end{equation}
In the thermodynamic $N\to \infty$ limit we find ${\cal A}_2= {\cal A}^2_{1}$ (i.e.~the vacuum contribution can be neglected) and
\begin{equation}\label{eq:GS_energy_thermodynamic}
\lim_{N\to \infty} \epsilon_{\rm GS} =- a\int_{-\pi/a}^{+\pi/a}\frac{dk}{2\pi} \epsilon(k) +\frac{\hbar\omega_{\rm c}}{4}{\cal A}^2_{1}+U \left( \frac{{\cal M}^2}{4} + |{\cal I}|^2 \right)~.
\end{equation}
Since $\epsilon(k)$ depends only on $ka$, it is useful to change integration variable in Eq.~(\ref{eq:GS_energy_thermodynamic}) from $k$ to $k'=ka \in (\pi,+\pi)$.

Eqs.~(\ref{gapeq-I}), (\ref{gapeq-M}), (\ref{gapeq2-J}), (\ref{gapeq2-T}), (\ref{min-A1}), and~(\ref{min-A2}) fully determine all the relevant quantities in the problem, i.e.~${\cal I}$, ${\cal M}$, ${\cal J}$, and ${\cal T}$.

\begin{figure}[t] 
\centering
\vspace{1.5em}
\begin{overpic}[width=0.40\columnwidth]{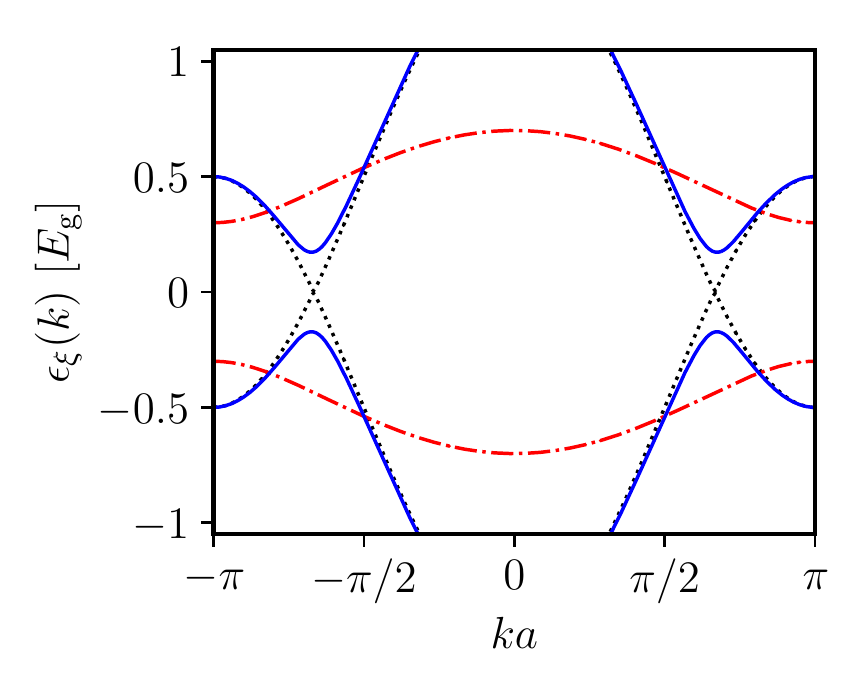}\put(4,80){\normalsize (a)}\end{overpic}
\begin{overpic}[width=0.40\columnwidth]{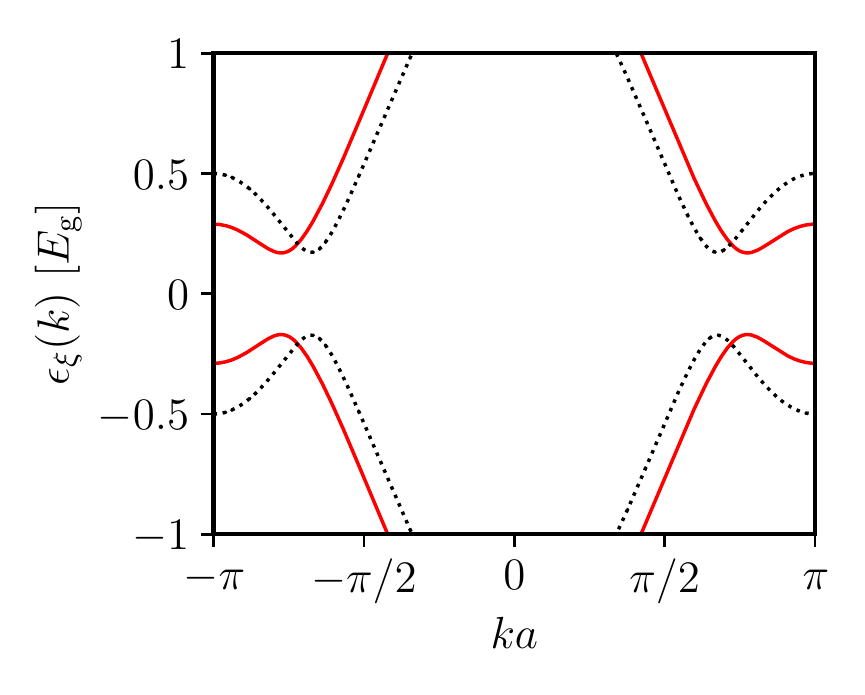}\put(4,80){\normalsize (b)}\end{overpic}
\caption{(Color online) Panel (a) The non-interacting $U=0$ spectrum  $\epsilon_{\xi}(k)=\xi\epsilon(k)=\pm \epsilon(k)$  (in units of $E_{\rm g}$) as a function of $ka$ in the first BZ, $ka\in(-\pi/\pi)$.  The red dashed line is the spectrum in an insulating non-interacting case $|t|<E_{\rm g}/4$, $t=0.1~E_{\rm g}$, and $\tilde{t}=0$. The black dotted line shows the metallic phase $|t|>E_{\rm g}/4$, $t=0.5~E_{\rm g}$, and $\tilde{t}=0$.
The blue solid line---obtained by setting $t=0.5~E_{\rm g}$ and $\tilde{t}=0.1~E_{\rm g}$---shows that a finite value of $\tilde{t}$ opens a single-particle hybridization gap. 
Panel (b) A comparison between the non-interacting and the interacting spectrum.  The black dotted line is the non-interacting spectrum (i.e.~obtained by setting $U=0$), for 
$t=0.5~E_{\rm g}$ and $\tilde{t}=0.1~E_{\rm g}$. The red solid line is the HF mean-field spectrum obtained for the same values of $t$ and $\tilde{t}$, at $U=E_{\rm g}$ (i.e.~$U/t=2$).\label{fig:S1}}
\end{figure}

In Fig.~\ref{fig:S1} we present a summary of our main results for the bands $\pm \epsilon(k)$ both in the simple non-interacting $U=0$ case---panel a)---and in the interacting $U\neq 0$ case---panel b).

\section{Section VII: Optical conductivity, Drude weight, and the $f$-sum rule}
\label{sect:optical-conductivity}

In this Section we discuss the optical conductivity $\sigma(\omega)$ and the $f$-sum rule for the EFK model. The longitudinal conductivity $\sigma(\omega)$ is defined as the response of the physical {\it charge} current operator to the electric field $E(t) = -c^{-1}\partial A(t)/\partial t$. Assuming $A(t) = A_{\omega}e^{-i\omega t}e^{\eta t}+{\rm c.c.}$ with $\eta =0^{+}$ (as usual, for the applicability of linear response theory the applied field must vanish in the far past~\cite{Giuliani_and_VignaleS,Pines_and_NozieresS}), we have $E(t) = i c^{-1} (\omega +i \eta) A_{\omega} e^{-i\omega t}e^{\eta t} + {\rm c.c.} = E_{\omega} e^{-i\omega t}e^{\eta t} + {\rm c.c.}$. 

We therefore find that the response of the physical current is given by
\begin{equation}\label{eq:general_conductivity}
-e \delta J_{\rm phys}(\omega) \equiv \sigma(\omega) E_{\omega} = \frac{i}{c} \sigma(\omega)(\omega +i \eta) A_{\omega}~.
\end{equation}
We conclude that the pre-factor in front of $A_{\omega}$ in the right-hand side Eq.~(\ref{eq:general_conductivity}) can be calculated from the current-current response function~\cite{Giuliani_and_Vignale,Pines_and_NozieresS}, with its paramagnetic and diamagnetic contributions.

Here, we focus on the EFK model, i.e.~Eq.~(9) for $g_{0}=0$. Using Eq.~(\ref{eq:general_conductivity}) and linear response theory~\cite{Giuliani_and_VignaleS,Pines_and_NozieresS}, we immediately find that the optical conductivity of the EFK model is given by
\begin{equation}\label{eq:exact_eigenstate_2}
\sigma(\omega) = \frac{i}{\omega  + i \eta}\frac{e^2a^2}{\hbar^2 L}\langle- \hat{\cal T}\rangle 
+\frac{e^2}{\hbar L}
\frac{i}{\omega+i\eta}\sum_{n,m}(P_{m}-P_{n})\frac{|\langle \psi_{n}|\hat{j}_{\rm p}|\psi_{m}\rangle|^2 }{\omega - \omega_{nm}+i \eta}~,
\end{equation}
where $|\psi_{n}\rangle$ are the exact eigenstates of the Hamiltonian $\hat{\cal H}_{0}+\hat{\cal H}_{\rm ee}$ with eigenvalues $E_{n}$, $\langle\dots\rangle \equiv \sum_{n}P_{n}\langle \psi_{n}|\dots|\psi_{n}\rangle$ denotes a thermal average,  and $P_{n} = \exp(-\beta E_{n})/{\cal Z}$, with $\beta = (k_{\rm B}T)^{-1}$ and ${\cal Z}=\sum_{n}\exp(-\beta E_{n})$ is the canonical partition function. In deriving the exact eigenstate representation (\ref{eq:exact_eigenstate_2}) we have used that $\hat{j}^\dagger_{\rm p} =  \hat{j}_{\rm p}$ and therefore $\langle \psi_{n}|\hat{j}_{\rm p}|\psi_{m}\rangle= \langle \psi_{m}|\hat{j}_{\rm p}|\psi_{n}\rangle^*$.

Separating the real and imaginary parts of $\sigma(\omega)$ and taking the zero-temperature limit ($P_{n}=0$ for $n\neq 0$ and $P_{0}=1$), we finally find:
\begin{equation}
{\rm Re}[\sigma(\omega)] = D \delta(\omega)  + \frac{\pi e^2}{L}  \sum_{n\neq 0}\frac{|\langle \psi_{n}|\hat{j}_{\rm p}|\psi_{0}\rangle|^2}{E_{n} - E_{0}} 
[\delta(\omega - \omega_{n0})+\delta(\omega + \omega_{n0})]
\end{equation}
where $D$ is the so-called Drude weight~\cite{kohn_pr_1964S,shastry_prl_1990S,millis_prb_1990S,fye_prb_1991S}
\begin{equation}\label{eq:Drude_latticemodel_zeroT}
D = \frac{\pi e^2a^2}{\hbar^2 L}\langle \psi_{0}|- \hat{\cal T}|\psi_{0}\rangle  - \frac{2\pi e^2}{L} \sum_{n\neq 0}\frac{|\langle \psi_{n}|\hat{j}_{\rm p}|\psi_{0}\rangle|^2 }{E_{n} - E_{0}}\equiv D_{\rm d}+{\rm D}_{\rm p}~.
\end{equation}
Here, $D_{\rm d}$ (${\rm D}_{\rm p}$) defines the diamagnetic (paramagnetic) contribution to $D$.

We immediately notice the $f$-sum rule~\cite{Giuliani_and_VignaleS,Pines_and_NozieresS,shastry_prl_1990S,millis_prb_1990S,fye_prb_1991S}
\begin{equation}\label{eq:zero-T-f-sum-rule}
\int_{-\infty}^{+\infty}d\omega {\rm Re}[\sigma(\omega)]=
2\int_{0}^{+\infty}d\omega {\rm Re}[\sigma(\omega)]=D + \frac{2 \pi e^2}{L}  \sum_{n\neq 0}\frac{|\langle \psi_{n}|\hat{j}_{\rm p}|\psi_{0}\rangle|^2 }{E_{n} - E_{0}} = D_{\rm d}~.
\end{equation}
\begin{figure}[h] 
\centering
\vspace{1.5em}
\begin{overpic}[width=0.40\columnwidth]{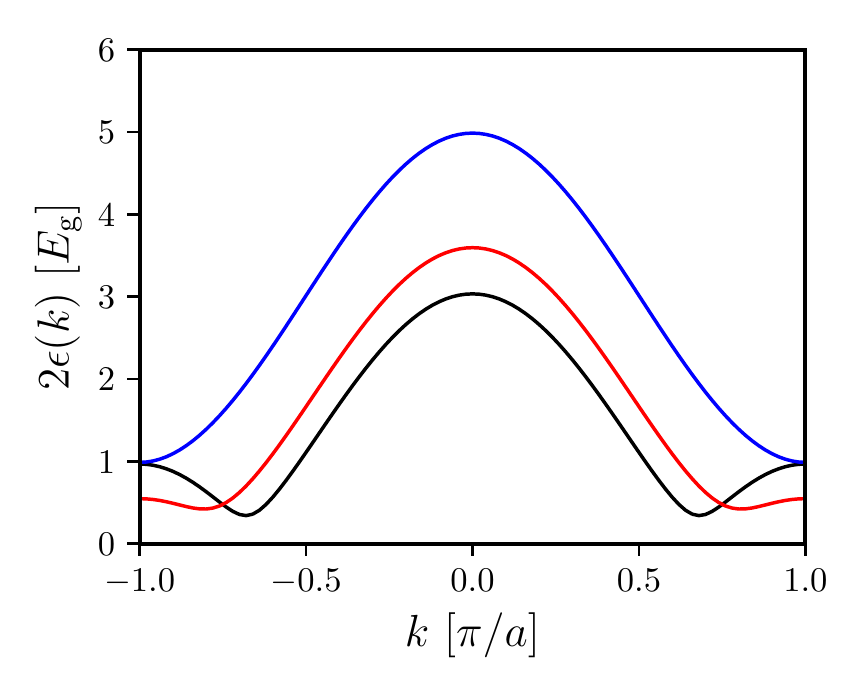}\put(4,80){(a)}\end{overpic}
\begin{overpic}[width=0.40\columnwidth]{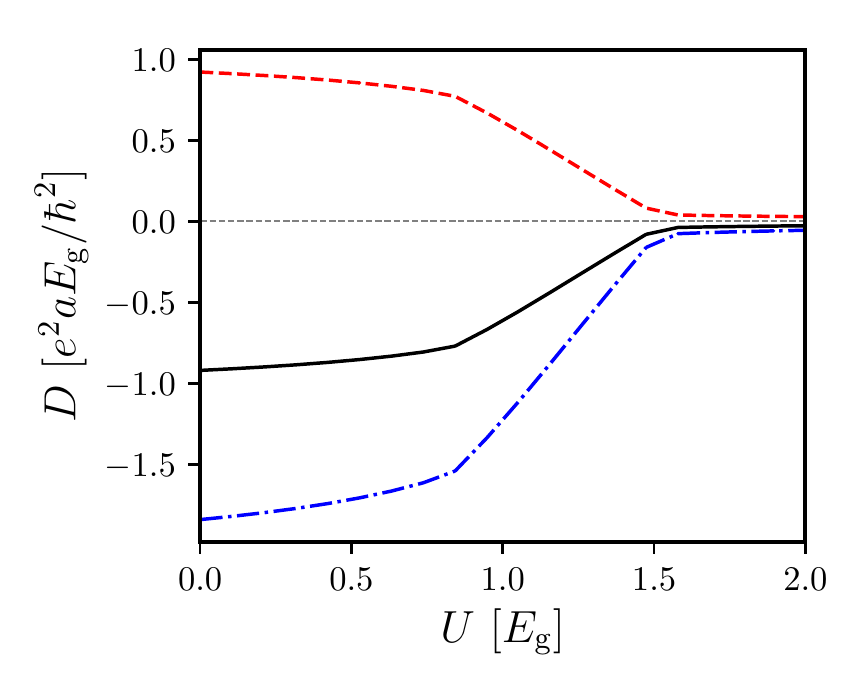}\put(4,80){(b)}\end{overpic}\\
\begin{overpic}[width=0.40\columnwidth]{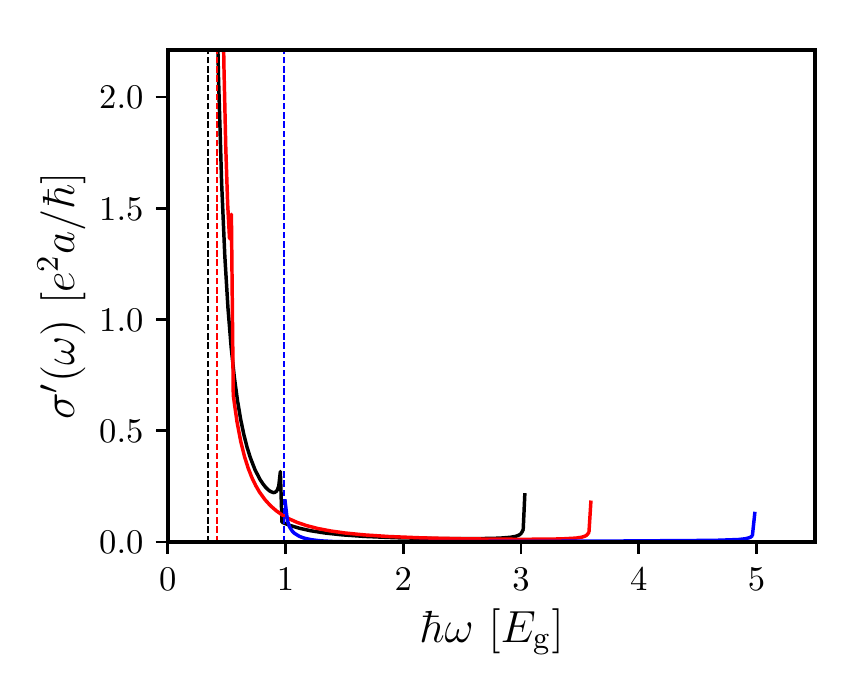}\put(4,80){(c)}\end{overpic}
\begin{overpic}[width=0.40\columnwidth]{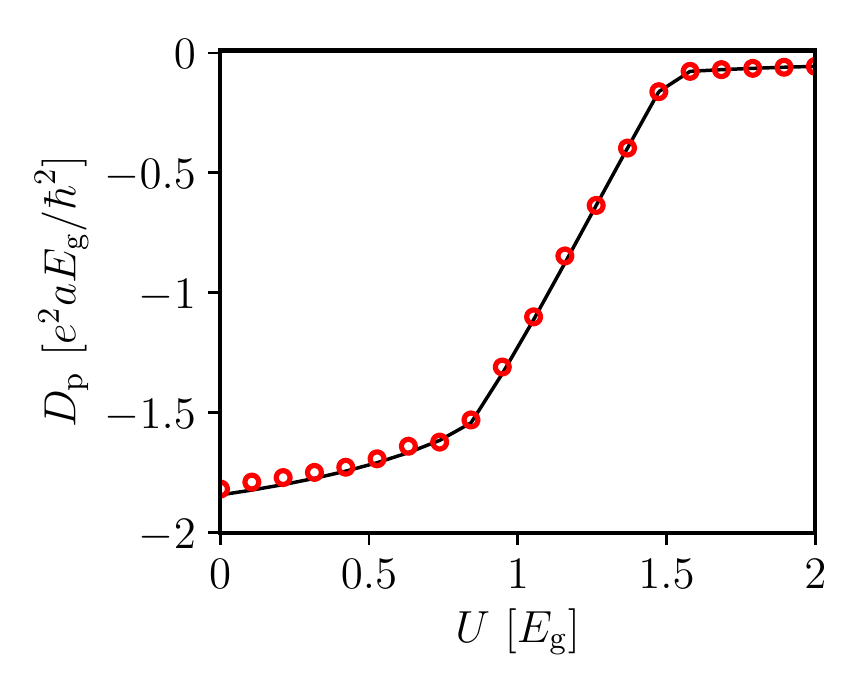}\put(4,80){(d)}\end{overpic}
\caption{(Color online) (a) The quantity  $2\epsilon(k)$ (in units of $E_{\rm g}$) as a function $ka$ in the first BZ. Different curves refer to different values of the Hubbard $U$ parameter: $U=E_{\rm g}/10$ (black), $U=E_{\rm g}$ (red), and $U=2E_{\rm g}$ (blue).
(b) The Drude weight $D$ (black solid line) and the contributions $D_{\rm p}$ (blue dash-dotted line) 
and $D_{\rm d}$ (red dashed line) are plotted as functions of $U/E_{\rm g}$. The three quantities $D$, $D_{\rm p}$, and $D_{\rm d}$ are in units of $e^2 a E_{\rm g}/\hbar^2$.
(c) The smooth contribution $\sigma'(\omega)$---see Eq.~(\ref{eq:smooth_sigma})---to the real part of the optical conductivity (in units of  $e^2a E_{\rm g}/\hbar$) is plotted as a function $\hbar \omega/E_{\rm g}$. Different curves refer to different values of the Hubbard $U$ parameter:
$U=E_{\rm g}/10$ (black solid line), $U=E_{\rm g}$ (red solid line), and $U=2E_{\rm g}$ (blue solid line). Black, red, and blue vertical dashed lines mark the energy $E_{\rm VHS}={\rm min}_{k\in {\rm BZ}}[2\epsilon(k)]$ at which a logarithmic divergence of $\sigma'(\omega)$ occurs. Clearly, $E_{\rm VHS}$ shifts with $U$.
(d) The paramagnetic contribution to the Drude weight $D_{\rm p}$ (black solid line, in units of $e^2 a E_{\rm g} /\hbar^2$) is compared with the quantity $S^\prime_{\rm p}$ (red circles) defined in Eq.~(\ref{eq:sprime}). All numerical results in this figure have been obtained by setting $t=E_{\rm g}/2$ and $\tilde{t}=E_{\rm g}/10$. 
\label{fig:fsr}}
\end{figure}

We now show that the $f$-sum rule is satisfied in our HF treatment of the EFK model. In the absence of light, the complete HF Hamiltonian including electron-electron interactions and neglecting an irrelevant constant is (see Eq.~(\ref{eq:Bogoliubov_Hamiltonian})):
\begin{eqnarray}\label{eq:Bogoliubov_Hamiltonian_no_light}
\hat{\cal H}^{\rm HF}=\sum_{k\in {\rm BZ}} \sum_{\xi=\pm} \xi \epsilon(k)\hat{\gamma}^\dagger_{k,\xi}\hat{\gamma}_{k,\xi}~.
\end{eqnarray}
The eigenstates and eigenvalues are $\ket{\xi,k} = \hat{\gamma}^\dagger_{\xi,k}\ket{{\rm vac}}$ and $\epsilon_{\xi}(k)=\xi\epsilon(k)$. We remind the reader that $\epsilon(k)$ has been defined in Eq.~(\ref{eq:spettro}) and, in this Section, needs to be evaluated at $g_{0}=0$. The ground state, as noticed above, is $\ket{\psi_{0}} = \prod_{k\in{\rm BZ}}\hat{\gamma}^\dagger_{-,k}\ket{{\rm vac}}$. We have
\begin{equation}\label{eq:Dp}
D_{\rm p}= - \frac{2 \pi e^2}{Na} \sum_{k\in{\rm BZ} }\frac{|\langle +,k|\hat{j}_{\rm p}|-,k\rangle|^2 }{2 \epsilon(k)}
\end{equation}
and
\begin{equation}
D_{\rm d}= \frac{\pi e^2 a}{\hbar^2 N} \sum_{k\in {\rm BZ}}  \langle -, k |- \hat{\cal T}|-, k \rangle~.
\end{equation}
The quantity $2\epsilon(k)$ is the energy necessary to promote an electron with wave number $k$ vertically from the lower band $\xi=-$ to the upper band $\xi=+$.
Fig.~\ref{fig:fsr}(a) shows the quantity $2\epsilon(k)$ (in units of $E_{\rm g}$) as a function of $ka$. The extrema of $2\epsilon(k)$ give rise to logarithmic divergences in the optical conductivity.

In order to calculate both contributions to the Drude weight, we need the following matrix elements:
\begin{equation}\label{eq:jpk}
\langle +, k | \hat{j}_{\rm p}|-, k\rangle = \frac{2a}{\hbar}
e^{i \phi_k}[t \sin(ka) \sin(\theta_k)+\tilde{t} \cos(ka) \cos(\theta_k) \sin(\phi_k) + i \tilde{t} \cos(ka) \cos(\phi_k)]
\end{equation}
and
\begin{equation}\label{eq:tk}
\langle -, k|- \hat{\cal T}|-,k\rangle =
[t \cos(ka) \cos(\theta_k)+\tilde{t} \sin(ka) \sin(\theta_k) \sin(\phi_k)]~,
\end{equation}
where $\theta_k$ and $\phi_k$ are the Bogoliubov angles defined in Sec.~VI. From Eq.~(\ref{eq:jpk}), we notice that, for $\tilde{t}=0$ and $U=0$, one has $\langle +, k | \hat{j}_{\rm p}|-,k\rangle=0$. This is expected since, in the absence of many-body effects and for $\tilde{t}=0$, the eigenstates have no orbital mixing (i.e.~$\sin(\theta_k)=0$). Switching on $\tilde{t}$ or $U$, however, yields $\langle +, k| \hat{j}_{\rm p}|-, k\rangle\neq 0$.
In particular, for $\tilde{t}=0$, repulsive interactions allow $\langle +, k| \hat{j}_{\rm p}|-,k\rangle\neq0$ when $\sin(\theta_k)\neq 0$, i.e.~for $0<U<U_{\rm XC} = U_{{\rm c}2}(0)$, since, from Eqs.~(\ref{Hk_1})-(\ref{Hk_3}), one has $\sin^2(\theta_k)=[U |{\cal I}|/\epsilon(k)]^2$.

Using Eqs.~(\ref{eq:jpk}) and~(\ref{eq:tk}), it is possible to calculate $D_{\rm p}$ and $D_{\rm d}$, and therefore $D$. Fig.~\ref{fig:fsr}(b) shows these quantities as functions of $U/E_{\rm g}$.

We now calculate the smooth part $\sigma^\prime(\omega)$ of ${\rm Re}[\sigma(\omega)]$,
\begin{equation}\label{eq:ReSgm}
\sigma^\prime(\omega) \equiv \frac{ \pi e^2}{Na} \sum_{k\in {\rm BZ}}  \frac{|\langle +, k| \hat{j}_{\rm p}|-,k\rangle|^2}{2 \epsilon(k)} 
[\delta(\omega - 2 \epsilon(k)/\hbar)+\delta(\omega + 2 \epsilon(k)/\hbar)]~.
\end{equation}
Note that $\sigma^\prime(\omega) = {\rm Re}[\sigma(\omega\neq0)]$.
Because of $\delta(\omega \mp 2 \epsilon(k)/\hbar)$ in the integrand of Eq.~(\ref{eq:ReSgm}), the integral over $k$ (in the thermodynamic limit) can be carried out analytically. We find
\begin{equation}\label{eq:smooth_sigma}
\sigma^\prime(\omega) = \frac{e^2 \hbar }{8} \sum_i  \frac{|\langle +, k| \hat{j}_{\rm p} |-, k\rangle|^2}{|\sum^3_{j=1} h_j(k) \partial_k h_j(k) |} \Bigg|_{k=k_i(\omega)}~,
\end{equation}
where $k_i(\omega)$ are the solutions of $\epsilon(k_i(\omega))=\hbar |\omega|/2$ and the quantities $h_1(k)$, $h_2(k)$, and $h_3(k)$ have been defined in Eqs.~(\ref{Hk_1})-(\ref{Hk_3}).
Fig.~\ref{fig:fsr}(c) shows $\sigma'(\omega)$ as a function $\hbar \omega/E_{\rm g}$. Each vertical dashed line marks the energy $E_{\rm VHS}=\min_{k\in {\rm BZ}}[2\epsilon(k)]$ at which the quantity $2\epsilon(k)$ is minimal. At this energy a logarithmic enhancement of $\sigma'(\omega)$ occurs. (Similarly, another singularity occurs at $E'_{\rm VHS}=\max_{k\in {\rm BZ}}[2\epsilon(k)]$, but that is weaker  in our numerical calculations.)

Using Eq.~(\ref{eq:smooth_sigma}) we can finally calculate numerically the quantity 
\begin{equation}\label{eq:sprime}
S^\prime_{\rm p} =-2\int_0^{+\infty} d\omega~ \sigma^\prime(\omega)~.
\end{equation}
We have verified numerically that 
\begin{equation}
S^\prime_{\rm p} =D_{\rm p}~.
\end{equation}
This is seen in Fig.~\ref{fig:fsr}(d). It follows that 
\begin{equation}
 2\int_0^{+\infty} d \omega {\rm Re} [\sigma(\omega)]=D_{\rm d}~,
\end{equation}
which is exactly the $f$-sum rule (\ref{eq:zero-T-f-sum-rule}).

\section{Section VIII: On the phase diagram of the EFK model}
\label{sect:Ucs}

In this Section we demonstrate the existence of two critical values of $U$, $U_{{\rm c}1}$ and $U_{{\rm c}2}$, at which ${\cal I}=0$, where ${\cal I}$ has been introduced in the self-consistent field equation (Eq.~(14)) of the main text and Eq.~(\ref{gapeq-I}). The latter yields the following equations:
\begin{equation}\label{eq:ReI}
\left[1- \frac{1}{N} \sum_{k\in {\rm BZ}} \frac{U}{2\epsilon(k)}\right]{\rm Re}({\cal I}) = 0
\end{equation}
and
\begin{equation}\label{eq:ImI}
\left[1- \frac{1}{N} \sum_{k\in {\rm BZ}}\frac{U}{2\epsilon(k)}\right]{\rm Im}({\cal I}) = -\tilde{t} \frac{1}{N} \sum_{k\in {\rm BZ}} \frac{\sin(ka) }{\epsilon(k)}
~. 
\end{equation}
In the absence of hybridization, i.e.~for $\tilde{t}=0$, the previous expressions become identical. This implies a degeneracy with to respect the phase of ${\cal I}$. 
For $\tilde{t}\neq 0$, these equations can be satisfied by solutions of the HF equations which yield ${\rm Im}({\cal I})=0$ and $\epsilon(-k)=\epsilon(k)$. The latter condition implies that the left-hand side of Eq.~(\ref{eq:ImI}) vanishes. All the solutions we find are of this type. 

The condition ${\rm Re}({\cal I})\neq 0$ implies that the following equation must be satisfied:
\begin{equation}\label{eq:EI_existence}
1- \frac{1}{N} \sum_{k\in {\rm BZ}} \frac{U}{2\epsilon(k)}=0~.
\end{equation}
Before proceeding to prove the existence of $U_{{\rm c}1}$ and $U_{{\rm c}2}$ we write Eq.~(\ref{eq:EI_existence}) in a more appealing form. We define $\psi_{k} = {\rm Re}({\cal I}) /[2 \epsilon(k)]$, and we rewrite Eq.~(\ref{eq:EI_existence}) as
\begin{equation}\label{eq:2particle}
2\epsilon(k)   \psi_{k} = \frac{1}{N}\sum_{k^\prime\in {\rm BZ}}  U \psi_{k^\prime}~.
\end{equation}
This establishes an immediate link between Eq.~(\ref{eq:EI_existence}) and the equation for the exciton binding energy.

We first demonstrate the existence of upper critical value of $U$, i.e.~$U_{{\rm c}2}(\tilde{t})$. Let us first set $\tilde{t}=0$. For $U/t\gg 1$, the system is in a trivial insulating phase in which all electrons occupy the ${\rm s}$ band and ${\cal M}={\cal M}_{0}\equiv -1$. Upon decreasing $U$ down to $U_{\rm XC} = U_{{\rm c}2}(0)$, the system develops an infinitesimal excitonic order parameter. The value $U_{\rm XC}$ at which this occurs can be found by solving Eq.~(\ref{eq:EI_existence}) for an infinitesimal ${\cal I}$, i.e.
\begin{equation}
 \frac{1}{N} \sum_{k\in{\rm BZ}} \frac{U_{\rm XC}}{4 t \cos(ka) + E_{\rm g}+U_{\rm XC}}=1~,
\end{equation}
or, in the thermodynamic limit,
\begin{equation}
\int_{-\pi}^\pi \frac{d x}{2\pi} \frac{1}{ \cos(x) + [E_{\rm g}+U_{\rm XC}]/(4t)}=\frac{4t}{U_{\rm XC}}~.
\end{equation}
Carrying out the integral analytically we find
\begin{equation}
U_{\rm XC}=\frac{8 t^2}{E_{\rm g}}- \frac{E_{\rm g}}{2}~.
\end{equation}
Corrections to $U_{\rm XC}$ and ${\cal M}_{0}$ due to $\tilde{t}\neq 0$ can be found perturbatively in the limit $\tilde{t}/t\ll 1$. They start at second order in the small parameter $\tilde{t}/t$: $\delta U_{{\rm c}2}(\tilde{t})= (\tilde{t}/t)^2 u$ and $\delta {\cal M}_0(\tilde{t})= (\tilde{t}/t)^2 m_0$. The latter is the change in the electronic polarization from the value ${\cal M}_{0}=-1$ in the limit $\tilde{t}=0$. We find $m_0 = E_{\rm g}/(2 U_{\rm XC})$ and $u = E^3_{\rm g}/[t(3 E^2_{\rm g} + 16 t^2 )]$. For example, for $t=E_{\rm g}/2$ we find $m_{0}=1/3$ and $u=2/7$. In conclusion, we have
\begin{equation}\label{eq:uc2}
U_{{\rm c}2}(\tilde{t}) = U_{\rm XC}+ 
\frac{E^3_{\rm g}(\tilde{t}/t)^2}{t(3 E^2_{\rm g} + 16 t^2 )}+{\cal O}(\tilde{t}^3)~.
\end{equation}

We now demonstrate the existence of a lower critical value of $U$, i.e.~$U_{{\rm c}1}(\tilde{t})$. Following similar steps to the ones above, one can demonstrate that there is also a lower-threshold for the existence of the exciton insulating phase. Up to leading order in an asymptotic expansion for small $\tilde{t}/t$ (and under the single-particle  condition $|t|>E_{\rm g}/4$ discussed in Fig.~\ref{fig:S1}) we find
\begin{equation}\label{eq:uc1}
U_{\rm c1}(\tilde{t}) \to \frac{ \pi \sqrt{4 t^2-E_{\rm g}^2/4} }{|\ln(\tilde{t}/t)|}~.
\end{equation}
We clearly see that $\lim_{\tilde{t}\to 0}U_{{\rm c}1}(\tilde{t})=0$. But, for finite $\tilde{t}/t$, $U_{\rm c1}(\tilde{t})\neq 0$. We will come back to Eq.~(\ref{eq:uc1}) below.

We have checked that the analytical results (\ref{eq:uc2}) and (\ref{eq:uc1}) match very well our numerical results, in their regime of validity.

\section{Section IX: Pseudospin analysis}
\label{sect:spins}

In this Section we present a few more remarks on the ground state of the EFK model in the HF approximation.

We view the mean-field problem as a variational problem and use a trial ground-state wave function of the form
\begin{equation}\label{eq:once_more}
| \psi \rangle = \prod_{k\in {\rm BZ}} \hat{\gamma}^\dagger_{k,-} | {\rm vac} \rangle~.
\end{equation}
We then express the full Hamiltonian of the 1D EFK model defined by Eq.~(\ref{eq:complete}) in terms of the Bogoliubov operators $\hat{\gamma}^\dagger_{k,\pm}$ and $\hat{\gamma}_{k,\pm}$:
\begin{eqnarray}
\hat{c}^\dagger_{k,{\rm s}}=u_{k} \hat{\gamma}^\dagger_{k,-} - v_{k} \hat{\gamma}^\dagger_{k,+}~,\\
\hat{c}^\dagger_{k,{\rm p}}=v^*_{k} \hat{\gamma}^\dagger_{k,-}+u^*_{k} \hat{\gamma}^\dagger_{k,+}~,
\end{eqnarray}
where $u_k=\cos(\theta_k/2)$ and $v_k=\sin(\theta_k/2)e^{i \phi_k}$.

By writing the Hamiltonian in its normal ordered form, exploiting the following property of the variational wave function
\begin{equation}
 \langle \psi |
 \hat{\gamma}^\dagger_{k_1,\lambda_1} \ldots \hat{\gamma}^\dagger_{k_n,\lambda_n}
 \hat{\gamma}_{k_{n+1},\lambda_{n+1}}\ldots \hat{\gamma}_{k_{2n}, \lambda_{2n} }
 | \psi \rangle
 =
 \prod_{j=1}^{2n} \delta_{\lambda_j,-}  
 \langle \psi |
 \hat{\gamma}^\dagger_{k_1,-} \ldots \hat{\gamma}^\dagger_{k_n,-}
 \hat{\gamma}_{k_{n+1},-}\ldots \hat{\gamma}_{k_{2n},-}
 | \psi \rangle 
 ~,
\end{equation}
and enforcing particle-hole symmetry, we find the following ground-state energy:
\begin{eqnarray}
&&\braket{\psi |\hat{\cal H} | \psi }
=\sum_{k\in {\rm BZ}} \left\{ [E_{\rm g}/2+2t\cos(ka)] \cos(\theta_k) +2\tilde{t}\sin(ka) \sin(\theta_k) \cos(\phi_k)
\right\}  \braket{\psi| \hat{\gamma}^\dagger_{k,-} \hat{\gamma}_{k,-} |\psi}  \nonumber \\
&+& \frac{U}{N} \sum_{k,k^\prime,q\in {\rm BZ}}  
 \sin(\theta_{k^\prime-q}/2) \cos(\theta_{k+q}/2) \cos(\theta_{k}/2)  \sin(\theta_{k^\prime}/2)   e^{i(\phi_{k^\prime}-\phi_{k^\prime-q})}
\braket{\psi|  \hat{\gamma}^\dagger_{k^\prime-q,-}\hat{\gamma}^\dagger_{k+q,-} \hat{\gamma}_{k,-} \hat{\gamma}_{k^\prime,-}|\psi}~.
\end{eqnarray}
Using the ansatz (\ref{eq:once_more}) in the previous equation and the properties
\begin{align}
 \braket{\psi|  \hat{\gamma}^\dagger_{k,-}\hat{\gamma}^\dagger_{k,-} |\psi}&=1~, \\
 \braket{\psi|  \hat{\gamma}^\dagger_{k^\prime-q,-}\hat{\gamma}^\dagger_{k+q,-} \hat{\gamma}_{k,-} \hat{\gamma}_{k^\prime,-}|\psi}&=
 \delta_{q,0} -  \delta_{q,k^\prime-k}~, 
\end{align}
we finally find
\begin{eqnarray}
\braket{\psi |\hat{\cal H} | \psi }&
=&\sum_{k\in {\rm BZ}} [E_{\rm g}/2+2t\cos(ka)] \cos(\theta_k) +2\tilde{t}\sin(ka) \sin(\theta_k) \cos(\phi_k)+
\frac{U}{N} \sum_{k,k^\prime \in {\rm BZ} } \cos^2(\theta_k/2) \sin^2(\theta_{k^\prime}/2) \nonumber \\
&-&  \frac{U}{N} \sum_{k,k^\prime \in {\rm BZ} } 
 \sin(\theta_{k}/2)  \cos(\theta_{k}/2)  \sin(\theta_{k^\prime}/2)  \cos(\theta_{k^\prime}/2) \cos(\phi_{k^\prime}-\phi_{k})~.
\end{eqnarray}
We therefore note that the ground-state energy can be written in a form that resembles the energy of a chain of classical interacting spins in an external magnetic field, i.e.
\begin{equation}
{\cal E}[ {\bm \tau}_k ]\equiv \braket{\psi | \hat{\cal H}| \psi} - \frac{UN}{4}=  -\sum_{k\in {\rm BZ}}  \left[B_y(k) \tau^y_k + B_z(k) \tau^z_k \right ] -\frac{U}{4N} \sum_{k,k^\prime \in {\rm BZ}} 
 {\bm \tau}_k \cdot {\bm \tau}_{k^\prime}~,
\end{equation}
where ${\bm B}(k)= [0,2\tilde{t}\sin(ka),E_{\rm g}/2+2t\cos(ka)]^{\rm T}$ and ${\bm \tau}_k$ is a unit vector with Cartesian components
$\tau^x_k \equiv \braket{\psi|\hat{c}_{{\rm p},k}^\dag \hat{c}_{{\rm s}, k} +\hat{c}_{{\rm s}, k}^\dag \hat{c}_{{\rm p}, k}  |\psi}=\sin(\theta_k) \cos(\phi_k)$, 
$\tau^y_k \equiv i \braket{\psi| \hat{c}_{{\rm p}, k}^\dag \hat{c}_{{\rm s}, k} - \hat{c}_{{\rm s}, k}^\dag \hat{c}_{{\rm p}, k}|\psi}=\sin(\theta_k) \sin(\phi_k)$, 
and $\tau^z_k = \braket{\psi|\hat{c}_{{\rm s},k}^\dag \hat{c}_{{\rm s}, k}  - \hat{c}_{{\rm p},k}^\dag \hat{c}_{{\rm p}, k} |\psi}=\cos(\theta_k)$. 
Notice that the ``lattice" of spins is in momentum rather than real space. 
In this pseudospin description, the repulsive Hubbard-$U$ interaction becomes a ferromagnetic rotationally-invariant spin-spin interaction term.
The spin configuration which minimizes the energy satisfies the self-consistent field equations
\begin{align}
\label{eq:meanfealdspin}
\tau^x_k &= -\frac{U}{2\mu_k N} \sum_{k\in {\rm BZ}}   \tau^x_k~,\\
\tau^y_k &= -\frac{B_y(k)}{\mu_k }-\frac{U}{2\mu_k N} \sum_{k\in {\rm BZ}}   \tau^y_k ~,\\
\tau^z_k &= -\frac{B_z(k)}{\mu_k } -\frac{U}{2 \mu_k N} \sum_{k\in {\rm BZ}}   \tau^z_k~,
\end{align}
where
\begin{equation}
\mu_k =  -\sqrt{ \left[ \frac{U}{2 N} \sum_{k\in {\rm BZ}}   \tau^x_k \right]^2+
\left[B_y(k)-\frac{U}{2 N} \sum_{k\in {\rm BZ}}   \tau^y_k \right]^2+
\left[B_z(k)-\frac{U}{2 N} \sum_{k\in {\rm BZ}}   \tau^z_k \right]^2}~. 
\end{equation}
The quantities ${\cal I}$ and ${\cal M}$ can also be expressed in terms of pseudospins:
\begin{equation}
 {\cal I} = \frac{1}{2N} \sum_{k\in {\rm BZ}}  (\tau^x_k-i\tau^y_k)
\end{equation}
and
\begin{equation}
 {\cal M} = -\frac{1}{N} \sum_{k\in {\rm BZ}}  \tau^z_k~.
\end{equation}

Within this description, we consider two limiting cases.
Firstly, we consider the limit $U \gg E_{\rm g},t, \tilde{t}$. Neglecting $E_{\rm g},t, \tilde{t}$ with respect to $U$, it follows that the following configurations
\begin{equation}\label{eq:taun}
 {\bm \tau}^{\rm (n)}_k=-\hat{\bm z}~,
\end{equation}
\begin{equation}\label{eq:taufx}
{\bm \tau}^{\rm (fx)}_k= \hat{\bm x}~,
\end{equation}
and
\begin{equation}\label{eq:taufy}
{\bm \tau}^{\rm (fy)}_k= \hat{\bm y}~,
\end{equation}
are degenerate (i.e. ${\cal E}[ {\bm \tau}^{\rm (n)}_k] ={\cal E}[ {\bm \tau}^{\rm (fx)}_k] ={\cal E}[ {\bm \tau}^{\rm (fy)}_k]=-UN/2$). 
For $U \gg E_{\rm g},t, \tilde{t}$ the system is invariant under rotations.
The configuration corresponding to ${\bm \tau}^{\rm (n)}_k$ describes the normal phase (${\cal I}=0$), while the ones corresponding to ${\bm \tau}^{\rm (fx)}_k$ and ${\bm \tau}^{\rm (fy)}_k$ correspond to HF states with ${\rm Re}({\cal I}) \neq0$ and ${\rm Im}({\cal I}) \neq0$, respectively.
This implies that all configurations of the form
\begin{equation}\label{eq:tautheta}
  {\bm \tau}^{(\theta,\phi)}_k= \cos(\theta)  {\bm \tau}^{\rm (n)}_k +  \sin(\theta)\cos(\phi) {\bm \tau}^{\rm (fx)}_k +\sin(\theta)\sin(\phi) {\bm \tau}^{\rm (fy)}_k
\end{equation}
are degenerate.
By turning on $E_{\rm g},t$ and $\tilde{t}$, and treating them as weak perturbations,
we find that the energy associated to the configurations (\ref{eq:tautheta}) is given by
${\cal E}[{\bm \tau}^{(\theta,\phi)}_k]=-[U+E_{\rm g} \cos(\theta)]N/2$. This means that the gap energy $E_{\rm g}$ makes the {\it normal} phase expressed in (\ref{eq:taun}) energetically preferred.
This simple example shows why at large values of the Hubbard-$U$ parameter, the HF phases with ${\cal I}\neq0$ do not occur.

Before concluding, we discuss a second limiting case.
We set $E_{\rm g}=0$ (which is compatible with the condition $|t|>E_{\rm g}/4$ described in Fig.~\ref{fig:S1}) and we assume $0<\tilde{t}<t$. 
Under these conditions, the external magnetic field ${\bm B}(k)$ lays on the $\hat{\bm y}$-$\hat{\bm z}$ plane and and its average value is zero, i.e.~$N^{-1} \sum_{k\in {\rm BZ}}  {\bm B}(k) ={\bm 0}$, 
but $B(k)\neq 0~\forall k$ if $t, \tilde{t}\neq0$.
The spin configuration which minimizes the energy is 
\begin{align}
\tau^x_k &= -\frac{U {\rm Re}({\cal I})}{\mu_k } ~,\\
\tau^y_k &= -\frac{2\tilde{t}\sin(ka)}{\mu_k } ~,\\
\tau^z_k &=-\frac{2 t \cos(ka)}{\mu_k } ~,\\
\mu_k &=  -\sqrt{ \left[ U {\rm Re}({\cal I}) \right]^2+
\left[2\tilde{t}\sin(ka)\right]^2+
\left[2 t \cos(ka) \right]^2}~,
\end{align}
with ${\rm Re}({\cal I})\neq 0$ only if the following implicit equation is satisfied:
\begin{equation}
\frac{U}{2N} \sum_{k\in {\rm BZ}}  \frac{1}{\sqrt{ \left[ U {\rm Re}({\cal I}) \right]^2+
\left[2\tilde{t}\sin(ka)\right]^2+
\left[2 t \cos(ka) \right]^2}} =1~.
\end{equation}
In the thermodynamic limit the latter can be rewritten as
\begin{equation}
 \frac{t}{U} = \frac{1}{ 2 \pi \sqrt{1+[U {\rm Re}({\cal I})]^2/(4t^2)} }
 {\rm K}
 \left( \sqrt{\frac{1-(\tilde{t}/t)^2}{1+[U {\rm Re}({\cal I})]^2/(4t^2)}}\right)~,
\end{equation}
where ${\rm K}(x)$ is the complete elliptic integral of the first kind. Inspecting
the right-hand side of the previous equation, one finds that, for fixed values of $t$, $\tilde{t}$, and $U$,
\begin{equation}
 0\le \frac{t}{U} \le  \frac{t}{U_{{\rm c}1}(\tilde{t})} \equiv \frac{1}{2\pi}  {\rm K} \left(\sqrt{1-(\tilde{t}/t)^2}\right)~,
\end{equation}
where $U_{{\rm c}1}$ is the minimum value of the Hubbard $U$ parameter which gives ${\rm Re}({\cal I})\neq 0$. For small values of $\tilde{t}$ we find
\begin{equation}\label{eq:from_the_elliptic_function}
U_{{\rm c}1}(\tilde{t}) \to \frac{2\pi t}{|\ln(\tilde{t}/t)|}~.
\end{equation}
Note the logarithmic divergence, as we has seen previously in Eq.~(\ref{eq:uc1}). The only role of $E_{\rm g}\neq 0$ is to replace
$2t \to \sqrt{4t^2-E_{\rm g}^2/4}$ in Eq.~(\ref{eq:from_the_elliptic_function}).

If $0<U<U_{{\rm c}1}$, the configuration which minimizes the energy is 
\begin{align}
\tau^x_k &= 0 ~,\\
\tau^y_k &= -\frac{2\tilde{t}\sin(ka)}{\mu_k } ~,\\
\tau^z_k &=-\frac{2 t \cos(ka)}{\mu_k } ~,\\
\mu_k &=  -\sqrt{ 
\left[2\tilde{t}\sin(ka)\right]^2+
\left[2 t \cos(ka) \right]^2}~,
\end{align}
which means that ${\rm Re}({\cal I})= 0$.
This second liming case well describes what occurs in the EFK model in the HF approximation for $U \sim U_{{\rm c}1}$.


\begin{thebibliography}{77}
%
\bibitem{dicke_pr_1954}
R.H. Dicke,  \href{https://doi.org/10.1103/PhysRev.93.99}{Phys. Rev.~{\bf 93}, 99 (1954)}.
%
\bibitem{gross_pr_1982}
M. Gross and S. Haroche, \href{https://doi.org/10.1016/0370-1573(82)90102-8}{Phys. Rep.~{\bf 93}, 301 (1982)}.
%
\bibitem{cong_josaB_2016}
K. Cong, Q. Zhang, Y. Wang, G.T. Noe II, A. Belyanin, and J. Kono, 
\href{https://doi.org/10.1364/JOSAB.33.000C80}{J. Opt. Soc. Am. B~{\bf 33}, C80 (2016)}.
%
\bibitem{kockum_naturereviewsphysics_2019}
A.F. Kockum, A. Miranowicz, S. De Liberato, S. Savasta, and F. Nori, 
\href{https://doi.org/10.1038/s42254-018-0006-2}{Nat. Rev. Phys.~{\bf 1}, 19 (2019)}.
%
%
\bibitem{kirton19}
P. Kirton, M.M. Roses, J. Keeling, and E.G. Dalla Torre, \href{https://doi.org/10.1103/10.1002/qute.201800043}{Adv. Quantum Technol.~{\bf 2}, 1970013 (2019)}.
%
\bibitem{noe_natphys_2012}
G.T. Noe II, J.-H. Kim, J. Lee, Y. Wang, A.K. W\'{o}jcik, S.A. McGill, D.H. Reitze, A.A. Belyanin, and J. Kono, \href{https://doi.org/10.1038/nphys2207}{Nat. Phys.~{\bf 8}, 219 (2012)}.
%
\bibitem{hepp_lieb}
K. Hepp and E.H. Lieb, \href{https://doi.org/10.1016/0003-4916(73)90039-0}{Ann. Phys.~{\bf 76}, 360 (1973)} and \href{https://doi.org/10.1103/PhysRevA.8.2517}{Phys. Rev. A~{\bf 8}, 2517 (1973)}.
%
\bibitem{wang_pra_1973}
Y.K. Wang and F.T. Hioe, \href{https://doi.org/10.1103/PhysRevA.7.831}{Phys. Rev. A~{\bf 7}, 831 (1973)}.
%
\bibitem{ETH}
See, however, experimental work on Dicke superradiance in open systems  formed by Bose-Einstein condensates coupled to optical cavities: K. Baumann, C. Guerlin, F. Brennecke, and T. Esslinger, \href{https://doi.org/10.1038/nature09009}{
Nature~{\bf 464}, 1301 (2010)}; K. Baumann, R. Mottl, F. Brennecke, and T. Esslinger, \href{https://doi.org/10.1103/PhysRevLett.107.140402}{Phys. Rev. Lett.~{\bf 107}, 140402 (2011)}; H. Ritsch, P. Domokos, F. Brennecke, and T. Esslinger, \href{https://doi.org/10.1103/RevModPhys.85.553}{Rev. Mod. Phys.~{\bf 85}, 553 (2013)}.
%
\bibitem{carmichael_physlett_1973}
H.J. Carmichael, C.W. Gardiner, and D.F. Walls, 
\href{https://doi.org/10.1016/0375-9601(73)90679-8}{Phys. Lett.~{\bf 46A}, 47 (1973)}.
%
\bibitem{rzazewski_prl_1975}
K. Rza\.{z}ewski, K. W\'{o}dkiewicz, and W. \.{Z}akowicz, 
\href{https://doi.org/10.1103/PhysRevLett.35.432}{Phys. Rev. Lett.~{\bf 35}, 432 (1975)};
I. Bialynicki-Birula and K. Rz\c{a}\.{z}ewski,
\href{https://doi.org/10.1103/PhysRevA.19.301}{Phys. Rev. A~{\bf 19}, 301 (1979)};
K. Gaw\c{e}dzki and K. Rz\c{a}\.{z}ewski,
\href{https://doi.org/10.1103/PhysRevA.23.2134}{Phys. Rev. A~{\bf 23}, 2134 (1981)}.
%
\bibitem{Sakurai}
J.J. Sakurai, {\it Modern Quantum Mechanics} (Addison-Wesley, Reading-MA, 1994).
%
\bibitem{Tufarelli15}
T. Tufarelli, K.R. McEnery, S.A. Maier, and M.S. Kim,
\href{https://doi.org/10.1103/PhysRevA.91.063840}{Phys. Rev. A~{\bf 91}, 063840 (2015)}.
%
\bibitem{emary_brandes}
C. Emary and T. Brandes, 
\href{https://doi.org/10.1103/PhysRevLett.90.044101}{Phys. Rev. Lett.~{\bf 90}, 044101 (2003)} and \href{https://doi.org/10.1103/PhysRevE.67.066203}{Phys. Rev. E~{\bf 67}, 066203 (2003)}; N. Lambert, C. Emary, and T. Brandes, \href{https://doi.org/10.1103/PhysRevLett.92.073602}{Phys. Rev. Lett.~{\bf 92}, 073602 (2004)}.
%
\bibitem{buzek_prl_2005}
V. Bu\v{z}ek, M. Orszag, and M. Ro\v{s}ko, 
\href{https://doi.org/10.1103/PhysRevLett.94.163601}{Phys. Rev. Lett.~{\bf 94}, 163601 (2005)}.
%
\bibitem{rzazewski_prl_2006}
K. Rz\c{a}\.{z}ewski and K. W\'{o}dkiewicz,
\href{https://doi.org/10.1103/PhysRevLett.96.089301}{Phys. Rev. Lett.~{\bf 96}, 089301 (2006)}.
%
\bibitem{nataf_naturecommun_2010}
P. Nataf and C. Ciuti, \href{https://doi.org/10.1038/ncomms1069 }{Nature Commun.~{\bf 1}, 72 (2010)}.
%
\bibitem{viehmann_prl_2011}
O. Viehmann, J. von Delft, and F. Marquardt,
 \href{https://doi.org/10.1103/PhysRevLett.107.113602}{Phys. Rev. Lett.~{\bf 107}, 113602 (2011)}.
%
\bibitem{ciuti_prl_2012}
C. Ciuti and P. Nataf, 
\href{https://doi.org/10.1103/PhysRevLett.109.179301}{Phys. Rev. Lett.~{\bf 109}, 179301 (2012)}.
%
\bibitem{Jaako16}
T. Jaako, Z.-L. Xiang, J.J. Garcia-Ripoll, and P. Rabl,
\href{https://doi.org/10.1103/PhysRevA.94.033850}{Phys. Rev.~A~{\bf 94}, 033850 (2016)}.
%
\bibitem{Bamba16}
M. Bamba, K. Inomata, and Y. Nakamura,
\href{https://doi.org/10.1103/PhysRevLett.117.173601}{Phys. Rev. Lett.~{\bf 117}, 173601 (2016)}.
%
\bibitem{hagenmuller_prl_2012}
D. Hagenm\"{u}ller and C. Ciuti, \href{https://doi.org/10.1103/PhysRevLett.109.267403}{Phys. Rev. Lett.~{\bf 109}, 267403 (2012)}.
%
\bibitem{chirolli_prl_2012}
L. Chirolli, M. Polini, V. Giovannetti, and A.H. MacDonald, 
\href{https://doi.org/10.1103/PhysRevLett.109.267404}{Phys. Rev. Lett.~{\bf 109}, 267404 (2012)}.
%
\bibitem{pellegrino_prb_2014}
F.M.D. Pellegrino, L. Chirolli, V. Giovannetti, R. Fazio, and M. Polini, \href{http://dx.doi.org/10.1103/PhysRevB.89.165406}{Phys. Rev. B {\bf 89}, 165406 (2014)}.
%
\bibitem{pellegrino_natcom_2016}
F.M.D. Pellegrino, V. Giovannetti, A.H. MacDonald, and M. Polini, 
\href{https://doi.org/10.1038/ncomms13355}{Nature Commun.~{\bf 7}, 13355 (2016)}.
%
\bibitem{scalari_science_2012}
G. Scalari, C. Maissen, D. Tur\v{c}inkov\'{a}, D. Hagenm\"{u}ller, S. De Liberato, C. Ciuti, C. Reichl, D. Schuh, 
W. Wegscheider, M. Beck, and J. Faist, \href{http://dx.doi.org/10.1126/science.1216022}{Science~{\bf 335}, 1323 (2012)}.
%
\bibitem{muravev_prb_2013}
V.M. Muravev, P.A. Gusikhin, I.V. Andreev, and I.V. Kukushkin, \href{https://doi.org/10.1103/PhysRevB.87.045307}{Phys. Rev. B~{\bf 87}, 045307 (2013)}.
%
\bibitem{smolka_science_2014}
S. Smolka, W. Wuester, F. Haupt, S. Faelt, W. Wegscheider, and A. Imamoglu,
\href{http://dx.doi.org/10.1103/10.1126/science.1258595}{Science~{\bf 346}, 332 (2014)}.
%
\bibitem{ravets_prl_2018}
S. Ravets, P. Kn\"{u}ppel, S. Faelt, M. Kroner, W. Wegscheider, and A. Imamoglu, \href{https://doi.org/10.1103/PhysRevLett.120.057401}{Phys. Rev. Lett.~{\bf 120}, 057401 (2018)}.
%
\bibitem{Paravicini-Bagliani_natphys_2019}
G.L. Paravicini-Bagliani, F. Appugliese, E. Richter, F. Valmorra, J. Keller, M. Beck, N. Bartolo, C. R\"{o}ssler, T. Ihn, K. Ensslin, C. Ciuti, G. Scalari, and J. Faist, \href{https://doi.org/10.1038/s41567-018-0346-y}{Nat. Phys.~{\bf 15}, 186 (2019)}. 
%
\bibitem{knuppel_arxiv_2019}
P. Kn\"{u}ppel, S. Ravets, M. Kroner, S. F\"{a}lt, W. Wegscheider, and A. Imamoglu, \href{https://arxiv.org/abs/1903.09256}{arXiv:1903.09256}.
%
\bibitem{EFKM}
For a recent review see e.g.~J. Kune\v{s},
\href{http://dx.doi.org/10.1088/0953-8984/27/33/333201}{J. Phys.: Condens. Matter~{\bf 27}, 333201 (2015)}.
%
\bibitem{Pines_and_Nozieres} 
D. Pines and P. Nozi\`eres, 
{\it The Theory of Quantum Liquids} (W.A. Benjamin, Inc., New York, 1966).
%
\bibitem{Giuliani_and_Vignale}
G.F. Giuliani and G. Vignale, {\it Quantum Theory of the Electron Liquid} (Cambridge University Press, Cambridge, 2005).
%
\bibitem{keeling_jpcm_2007}
J. Keeling, \href{https://doi.org/10.1088/0953-8984/19/29/295213}{J. Phys.: Condens. Matter~{\bf 19}, 295213 (2007)}.
%
\bibitem{Vukics12}
A. Vukics and P. Domokos,
\href{https://doi.org/10.1103/PhysRevA.86.053807}{Phys. Rev. A~{\bf 86}, 053807 (2012)}.
%
\bibitem{Stokes19}
A. Stokes and A. Nazir,
\href{https://arxiv.org/abs/1905.10697}{arXiv:1905.10697}.
%
\bibitem{SOM}
See the Supplemental Material file for  technical details on the ground-state factorization, on the stiffness theorem, on the TRK sum rule, on the coupling of light to the EFK model degrees of freedom, on the HF treatment of electron-electron interactions and the resulting Bogoliubov transformation, on the $f$-sum rule, and on the phase diagram of the EFK model.
%
\bibitem{Walls_and_Milburn}
D.F. Walls and G.J. Milburn, 
{\it Quantum Optics} (Springer, Berlin, 2007).
%
\bibitem{Serafini} 
A. Serafini, {\it Quantum Continuous Variables: A Primer of Theoretical Methods} (CRC Press, Boca Raton, FL, 2017).
%
\bibitem{first-order}
In the thermodynamic limit, the ground-state factorization implies a photon coherent state. The matter state for a given $\beta$ is therefore equivalent to a
matter state with a time-independent spatially constant vector potential. In turn, this implies a matter state energy independent of $\beta$. According to Eq.~(\ref{eqHL2}), introducing photons only costs energy, forbidding a first-order transition to a photon condensate.
%
\bibitem{excitonic_insulators}
See, for example, D. J\'{e}rome, T.M. Rice, and W. Kohn, \href{https://doi.org/10.1103/PhysRev.158.462}{Phys. Rev.~{\bf 158}, 462 (1967)};
B.I. Halperin and T.M. Rice, \href{https://doi.org/10.1016/S0081-1947(08)60740-7}{Solid State Phys.~{\bf 21}, 115 (1968)}; 
B.I. Halperin and T.M. Rice, \href{https://doi.org/10.1103/RevModPhys.40.755}{Rev. Mod. Phys.~{\bf 40}, 755 (1968)}.
%
\bibitem{portengen_prl_1996}
T. Portengen, Th. \"{O}streich, and L.J. Sham, \href{https://doi.org/10.1103/PhysRevLett.76.3384}{Phys. Rev. Lett.~{\bf 76}, 3384 (1996)} and 
\href{https://doi.org/10.1103/PhysRevB.54.17452}{Phys. Rev. B~{\bf 54}, 17452 (1996)}.
%
\bibitem{batista_prl_2002}
C.D. Batista, 
\href{https://doi.org/10.1103/PhysRevLett.89.166403}{Phys. Rev. Lett.~{\bf 89}, 166403 (2002)}.
%
\bibitem{kohn_pr_1964}
W. Kohn, \href{https://doi.org/10.1103/PhysRev.133.A171}{Phys. Rev.~{\bf 133}, A171 (1964)}.
%
\bibitem{shastry_prl_1990}
B.S. Shastry and B. Sutherland, \href{https://doi.org/10.1103/PhysRevLett.65.243}{Phys. Rev. Lett.~{\bf 65}, 243 (1990)}.
%
\bibitem{millis_prb_1990}
A.J. Millis and S.N. Coppersmith, \href{https://doi.org/10.1103/PhysRevB.42.10807}{Phys. Rev. B~{\bf 42}, 10807(R) (1990)}.
%
\bibitem{fye_prb_1991}
R.M. Fye, M.J. Martins, D.J. Scalapino, J. Wagner, and W. Hanke, \href{https://doi.org/10.1103/PhysRevB.44.6909}{Phys. Rev. B~{\bf 44}, 6909 (1991)}.
%
\bibitem{f-sum-rule}
Gauge invariance (of linear-response functions) and the $f$-sum rule follow from particle (charge) conservation~\cite{Giuliani_and_Vignale,Pines_and_Nozieres}. Consider Eq.~(\ref{eq:quadratic_Hamiltonian_final}) at $g_{0}=0$ and arbitrary values of $U$. In this case, the conservation law reads as following, $\partial_t \hat{n}_{\ell}+(\hat{j}_{{\rm p}, \ell+1}-\hat{j}_{{\rm p},\ell})/a=0$, which is the lattice version of 1D continuity equation. Here, we have defined the local density operator in the site representation, $\hat{n}_{\ell}\equiv\sum_{\alpha}\hat{c}^\dagger_{\ell,\alpha}\hat{c}_{\ell,\alpha}$ and the local paramagnetic (number) current operator 
$\hat{j}_{{\rm p},\ell} \equiv it_{\rm s} a
(\hat{c}^\dagger_{\ell, {\rm s}}\hat{c}_{\ell-1,  {\rm s}}-\hat{c}^\dagger_{\ell-1,  {\rm s}}\hat{c}_{\ell, {\rm s}})/\hbar
-it_{\rm p} a(\hat{c}^\dagger_{\ell, {\rm p}}\hat{c}_{\ell-1,  {\rm p }}-\hat{c}^\dagger_{\ell-1,  {\rm p }}\hat{c}_{\ell, {\rm p}} )/\hbar+i\tilde{t} a(\hat{c}^\dagger_{\ell, {\rm s}}\hat{c}_{\ell-1,  {\rm p}}  - \hat{c}^\dagger_{\ell-1,  {\rm p}}\hat{c}_{\ell, {\rm s}})/\hbar+ i\tilde{t} a(\hat{c}^\dagger_{\ell-1, {\rm s}}\hat{c}_{\ell,  {\rm p }}- \hat{c}^\dagger_{\ell,  {\rm p }}\hat{c}_{\ell-1, {\rm s}})/\hbar$. Coupling to the uniform vector potential of the cavity {\it must} be done via the paramagnetic current operator $\hat{j}_{\rm p}=\sum_{\ell=1}^{N}\hat{j}_{{\rm p},\ell}$ (while, at the same time, including the diamagnetic term). This is manifestly displayed by our Hamiltonian (\ref{eq:quadratic_Hamiltonian_final}) at $g_{0}\neq 0$.
Further details on the $f$-sum rule can be found in Sec.~VII of the SM~\cite{SOM}.
%
\bibitem{distefano_arXiv_2018}
O. Di Stefano, A. Settineri, V. Macr\`{i}, L. Garziano, R. Stassi, S. Savasta, and F. Nori, \href{https://doi.org/10.1038/s41567-019-0534-4}{Nat. Phys.~(2019)}.
%
\bibitem{DeBernardis18}
D. De Bernardis, T. Jaako, and P. Rabl,
\href{https://doi.org/10.1103/PhysRevA.97.043820}{Phys. Rev. A~{\bf 97}, 043820 (2018)}.
%
\bibitem{DeBernardis18b}
D. De Bernardis, P. Pilar, T. Jaako, S. De Liberato, and P. Rabl,
\href{https://doi.org/10.1103/PhysRevA.98.053819}{Phys. Rev. A~{\bf 98}, 053819 (2018)}.
%
\bibitem{verges}
J.A. Verg\'es, E. Louis, P.S. Lomdahl, F. Guinea, and A.R. Bishop, \href{https://doi.org/10.1103/PhysRevB.43.6099}{Phys. Rev. B~{\bf 43}, 6099 (1991)}; J.A. Verg\'es,  F. Guinea, and E. Louis, \href{https://doi.org/10.1103/PhysRevB.46.3562}{Phys. Rev. B~{\bf 46}, 3562 (1992)}.
%
\bibitem{kocharian_prb_1996}
A.N. Kocharian and J.H. Sebold, \href{https://doi.org/10.1103/PhysRevB.53.12804}{Phys. Rev. B~{\bf 53}, 12804 (1996)}.
%
\bibitem{ejima_prl_2014}
S. Ejima, T. Kaneko, Y. Ohta, and H. Fehske, \href{https://doi.org/10.1103/PhysRevLett.112.026401}{Phys. Rev. Lett.~{\bf 112}, 026401 (2014)}.
%
\bibitem{mazza_prl_2019}
The influence of strong interactions combined with a uniform single-mode cavity field on the phase diagram of small-gap semiconductors has also been considered in an interesting 
recent paper by G. Mazza and A. Georges, \href{https://doi.org/10.1103/PhysRevLett.122.017401}{Phys. Rev. Lett.~{\bf 122}, 017401 (2019)}. Some of the conclusions of this paper are incorrect because the model is not precisely gauge invariant.
%
\bibitem{Schlawin_prl_2019}
F. Schlawin, A. Cavalleri, and D. Jaksch, \href{https://doi.org/10.1103/PhysRevLett.122.133602}{Phys. Rev. Lett.~{\bf 122}, 133602 (2019)}.
%
\bibitem{curtis_prl_2019}
J.B. Curtis, Z.M. Raines, A.A. Allocca, M. Hafezi, and V.M. Galitski, \href{https://doi.org/10.1103/PhysRevLett.122.167002}{Phys. Rev. Lett.~{\bf 122}, 167002 (2019)}.
%
\bibitem{GMP}
S.M. Girvin, A.H. MacDonald, and P.M. Platzman, \href{https://doi.org/10.1103/PhysRevLett.54.581}{Phys. Rev. Lett.~{\bf 54}, 581 (1985)} and \href{https://doi.org/10.1103/PhysRevB.33.2481}{Phys. Rev. B~{\bf 33}, 2481 (1986)}; S.M. Girvin and A.H. MacDonald, \href{https://doi.org/10.1103/PhysRevLett.58.1252}{Phys. Rev. Lett.~{\bf 58}, 1252 (1987)}; F.D.M. Haldane, \href{https://doi.org/10.1103/PhysRevLett.107.116801}{Phys. Rev. Lett.~{\bf 107}, 116801 (2011)}.
%
\end{thebibliography}

\begin{thebibliography}{77}
%
\bibitem{Giuliani_and_VignaleS}
G.F. Giuliani and G. Vignale, {\it Quantum Theory of the Electron Liquid} (Cambridge University Press, Cambridge, 2005).
%
\bibitem{Pines_and_NozieresS} 
D. Pines and P. Nozi\`eres, 
{\it The Theory of Quantum Liquids} (W.A. Benjamin, Inc., New York, 1966).
%
\bibitem{kohn_pr_1964S}
W. Kohn, \href{https://doi.org/10.1103/PhysRev.133.A171}{Phys. Rev.~{\bf 133}, A171 (1964)}.
%
\bibitem{shastry_prl_1990S}
B.S. Shastry and B. Sutherland, \href{https://doi.org/10.1103/PhysRevLett.65.243}{Phys. Rev. Lett.~{\bf 65}, 243 (1990)}.
%
\bibitem{millis_prb_1990S}
A.J. Millis and S.N. Coppersmith, \href{https://doi.org/10.1103/PhysRevB.42.10807}{Phys. Rev. B~{\bf 42}, 10807(R) (1990)}.
%
\bibitem{fye_prb_1991S}
R.M. Fye, M.J. Martins, D.J. Scalapino, J. Wagner, and W. Hanke, \href{https://doi.org/10.1103/PhysRevB.44.6909}{Phys. Rev. B~{\bf 44}, 6909 (1991)}.
%
\end{thebibliography}
\end{document}